\documentclass[11pt, a4paper]{article}
\usepackage{amsmath}

\newtheorem{theorem}{Theorem}

\newcommand{\benumerate}{\begin{enumerate}}
\newcommand{\eenumerate}{\end{enumerate}}

\newcommand{\bitemize}{\begin{itemize}}

\newcommand{\eitemize}{\end{itemize}}

\begin{document}

\title{Hamiltonian systems of hydrodynamic type in $2+1$ dimensions}
\author{E.V. Ferapontov$^1$, A. Moro$^1$ and V.V. Sokolov$^2$}
    \date{}
    \maketitle
    \vspace{-7mm}
\begin{center}
$^1$ Department of Mathematical Sciences \\ Loughborough University \\
Loughborough, Leicestershire LE11 3TU \\ United Kingdom \\[2ex]
$^2$ Landau Institute for Theoretical Physics \\ Kosygina 2, 119334 Moscow, Russia \\[2ex]
e-mails: \\[1ex] \texttt{E.V.Ferapontov@lboro.ac.uk}\\
\texttt{A.Moro@lboro.ac.uk}\\
\texttt{vsokolov@landau.ac.ru}
\end{center}

\bigskip

\begin{abstract}

We investigate multi-dimensional Hamiltonian systems  associated with constant Poisson brackets
of hydrodynamic type. A complete list of two- and three-component  integrable Hamiltonians is obtained.
All our examples possess dispersionless Lax pairs and  an infinity of hydrodynamic reductions.

\bigskip

\noindent MSC: 35L40, 35L65, 37K10.

\bigskip

Keywords: Multi-dimensional Systems of  Hydrodynamic Type, Hamiltonian Structures, Dispersionless Lax Pairs, Hydrodynamic Reductions.
\end{abstract}

\newpage

\section{Introduction}

Over the past three decades there has been a significant progress in the theory of  $(1+1)$-dimensional
quasilinear  systems,
\begin{equation}
u^i_t+v^i_j({\bf u}) u^j_x=0,
\label{1+1}
\end{equation}
which are representable in the Hamiltonian form
$
u^i_t+P^{ij}h_j=0.
$
Here $h({\bf u})$ is a Hamiltonian density, $h_j=\partial_{u^j}h$, and  $P^{ij}$ is a  Hamiltonian operator of  differential-geometric type,
$$
P^{ij}=g^{ij}({\bf u}) \frac{d}{dx}+b^{ij}_k({\bf u})u^k_x,
$$
generated by a metric $g^{ij}$ (assumed non-degenerate) and its Levi-Civita connection
$\Gamma^i_{jk}$ via $b^{ij}_k=-g^{is}\Gamma^j_{sk}$.  It was demonstrated in \cite{Dubrovin1} that
the metric  $g^{ij}$ must necessarily be flat,  and in the flat coordinates of $g^{ij}$ the operator $P^{ij}$
takes a constant coefficient form $P^{ij}=\epsilon^i\delta^{ij}\frac{d}{dx}$. In the same coordinates,
Hamiltonian systems take a Hessian form $u^i_t+\epsilon^i h_{ij} u^j_x=0.$ It was observed that many
particularly important examples  arising in applications are diagonalizable, that is, reducible to the
Riemann invariant form $R^i_t+v^i({\bf R}) R^i_x=0.$ We recall that there exists a simple tensor
criterion of the diagonalizability for an arbitrary hyperbolic system (\ref{1+1}).
Let us first calculate the Nijenhuis tensor of the matrix $v^i_j$,
\begin{equation}
{\cal
N}^i_{jk}=v^p_j\partial_{u^p}v^i_k-v^p_k\partial_{u^p}v^i_j-v^i_p(\partial_{u^j}v^p_k-\partial_{u^k}v^p_j),
\label{N}
\end{equation}
 and introduce the Haantjes tensor
\begin{equation}
{\cal H}^i_{jk}={\cal N}^i_{pr}v^p_jv^r_k-{\cal
N}^p_{jr}v^i_pv^r_k-{\cal N}^p_{rk}v^i_pv^r_j+{\cal
N}^p_{jk}v^i_rv^r_p. \label{Ha}
\end{equation}
It was observed in \cite{Haantjes}
that a  $(1, 1)$-tensor  $v^i_j$ with mutually distinct eigenvalues is diagonalizable if and only if the
corresponding  Haantjes tensor $\cal H$ is identically zero.
As demonstrated by Tsarev,  a combination of the diagonalizability with the Hamiltonian property implies
the integrability: all diagonalizable Hamiltonian systems possess an infinity of conservation laws and
commuting flows, and can be solved by the generalized hodograph transform. We refer to \cite{Tsarev, Dubrovin1}
for further discussion and references.

The aim of our paper is to generalize this approach to $(2+1)$-dimensional Hamiltonian systems
\begin{equation}
{\bf u}_t+A({\bf u}){\bf u}_x+B({\bf u}){\bf u}_y=0,
\label{2+1}
\end{equation}
which are representable  in the form
$
{\bf u}_t+P h_{\bf u}=0
$
where  $h({\bf u})$ is a Hamiltonian density, and $P$ is a two-dimensional Hamiltonian operator of  differential-geometric type,
$$
P^{ij}=g^{ij}({\bf u}) \frac{d}{dx}+b^{ij}_k({\bf u})u^k_x+\tilde g^{ij}({\bf u}) \frac{d}{dy}+
\tilde b^{ij}_k({\bf u})u^k_y;
$$
such operators are generated by a pair of metrics $g^{ij}, \ \tilde g^{ij}$ and the corresponding Levi-Civita
connections $\Gamma^i_{jk}, \tilde \Gamma^i_{jk}$ via $b^{ij}_k=-g^{is}\Gamma^j_{sk}, \
\tilde b^{ij}_k=-\tilde g^{is}\tilde \Gamma^j_{sk}$. The theory of multi-dimensional Poisson brackets was
constructed in \cite{Dubrovin1, Mokhov1, Mokhov2}. The main difference from the one-dimensional situation
is that, although both metrics $g^{ij}$ and $ \tilde g^{ij}$ must necessarily be flat, they can no longer
be reduced to a constant coefficient form simultaneously: there exist obstruction tensors. The obstruction
tensors are known to vanish if either one of the metrics is positive definite, or a pair of metrics
 is non-singular in the sense of \cite{Mokhov2}, that is, the mutual eigenvalues of $g^{ij}$ and $ \tilde g^{ij}$ are
 distinct. In both cases, the operator $P^{ij}$ can be transformed to a constant coefficient form.
In the two-component situation any non-singular Hamiltonian operator can be cast into a canonical form
$$
P= \left(
\begin{array}{cc}
d/dx&0\\
0 & d/dy
\end{array}
\right)
$$
by an appropriate linear change of the independent variables $x, y$. The corresponding Hamiltonian systems
take the form
\begin{equation}
u^1_t+(h_1)_x=0, ~~~ u^2_t+(h_2)_y=0.
\label{2comp}
\end{equation}
The `simplest'  non-trivial  integrable Hamiltonian density  is
$h(u^1, u^2)=u^1u^2-\frac{1}{6}(u^1)^3$ (we point out that, up to
certain natural equivalence, there exist no other integrable
densities which are polynomial in $u^1, u^2$). The corresponding
equations (\ref{2comp}) take the form
$$
u^1_t-u^1u^1_x+u^2_x=0, ~~~ u^2_t+u^1_y=0,
$$
see Sect. 4.1. This system appears in the context of the genus zero universal Whitham hierarchy, \cite{Kr, Manas}.
Setting $u^1=-\varphi_{xt}, \ u^2=\varphi_{xy}$ one obtains a second order PDE
$$
\varphi_{tt}-\varphi_{xy}+\frac{1}{2}\varphi^2_{xt}=0,
$$
which is one of the   Hirota equations of the dispersionless Toda hierarchy \cite{F}.
The same equation appeared  in \cite{Pav} in the classification of integrable Egorov's
hydrodynamic chains. Other examples of integrable Hamiltonian densities  expressible in
elementary functions include
$$
h(u^1, u^2)=\frac{1}{2}(u^1-u^2)^2+e^{u^2}, ~~~ h(u^1, u^2)=u^2\sqrt {u^1}+\alpha (u^1)^{5/2},
~~~ h(u^1, u^2)=(u^1u^2)^{2/3},
$$
etc. The problem of classification of integrable two-component Hamiltonian systems (\ref{2comp})
was first addressed in \cite{Fer1} based on the method of hydrodynamic reductions.
We recall that a multi-dimensional quasilinear system (\ref{2+1}) is said to be integrable
if it possesses an infinity of $n$-component hydrodynamic reductions parametrized by $n$ arbitrary
functions of a single variable (see Sect. 2 for more details). It was demonstrated in \cite{Fer1} that this requirement imposes strong
restrictions on the corresponding Hamiltonian density $h(u^1, u^2)$. In Sect. 4 we  provide a complete list of integrable Hamiltonian densities (Theorem 1),
as well as the associated dispersionless Lax pairs (Sect. 4.1). The `generic' density is expressed
in terms of the Weierstrass elliptic functions.

In the three-component situation we consider Hamiltonian operators of the form
\begin{equation}
P=\left(\begin{array}{ccc}
{1} & 0 & 0\\
0 & {1} & 0\\
0&0&{1}
\end{array}
\right)\frac{d}{dx}+
\left(\begin{array}{ccc}
\lambda^{1} & 0 & 0\\
0 & \lambda^{2} & 0\\
0&0&\lambda^{3}
\end{array}
\right)\frac{d}{dy},
\label{3Ham}
\end{equation}
here $\lambda^i$ are constant and pairwise distinct; the corresponding Hamiltonian systems are
\begin{equation}
u^i_t+(h_i)_x+\lambda^i(h_i)_y=0.
\label{Ham}
\end{equation}
There is a new phenomenon arising in the  multi-component case: it
was observed in \cite{Fer2} that the necessary condition for
integrability of an $n$-component quasilinear system (\ref{2+1}) is the vanishing
of the Haantjes tensor for an arbitrary matrix of the form
$$
(\alpha A +\beta B+\gamma I_{n})^{-1}(\tilde \alpha A +\tilde
\beta B+\tilde \gamma I_{n}).
$$
In fact, it is sufficient to require the vanishing of the Haantjes
tensor for a two-parameter family $(kA + I_{n})^{-1}( lB+ I_{n})$.
We point out that  in the two-component case the Haantjes tensor
vanishes automatically. On the contrary, in the multi-component
situation the vanishing of the Haantjes tensor is a very strong
restriction. Systems with this property will be called
`diagonalizable' (we would like to stress that matrices $A$ and
$B$ do not commute in general, and cannot be diagonalized
simultaneously). In Sect. 5 we obtain a complete list of
diagonalizable three-component Hamiltonian systems (\ref{Ham})
(Theorem 3). It turns out that in this case the diagonalizability
conditions are very restrictive, and imply the integrability. For
technical reasons, the classification results take much simpler
form when expressed in terms of the Legendre transform $H$ of the
Hamiltonian density $h$, rather then the Hamiltonian density $h$
itself (recall that $H = \sum u^ih_i-h,  \ H_i=u^i, \ u_i=h_i$; we
use variables $u_i$ with lower indices for the arguments of $H$).
We demonstrate that the Legendre transform $H$ of the `generic'
integrable Hamiltonian density  $h$ is given by the formula
$$
 H=\sum_{j\ne i}\frac{\lambda^i-\lambda^j}{a^2_ia^2_j}V(a_iu_i, a_ju_j)
$$
where
$$
V(x, y)=Z (x-y)+\epsilon Z (x-\epsilon y)+
\epsilon^2 Z (x-\epsilon^2 y);
$$
here $a_i$ are arbitrary constants,  $\epsilon=e^{2\pi i/3}$, and $Z''=\zeta$ where $\zeta$ is the Weierstrass zeta-function: $\zeta'=-\wp, \  (\wp')^2=4\wp^3-g_3$. Notice that we are dealing with an incomplete elliptic curve, $g_2=0$, and that the expression for  $V$ is real. The above formula for $H$ has a natural  multi-component extension, which is also  integrable. This formula possesses a number of remarkable degenerations which are listed in  Theorems 1 and 3. In particular, one has
$$
H=\sum_{j\ne i}\frac{\lambda^i-\lambda^j}{a^2_ia^2_j}(a_iu_i-a_ju_j)\ln (a_iu_i-a_ju_j).
$$
We prove that all examples appearing in the classification possess dispersionless Lax pairs and an infinity of hydrodynamic reductions (Theorems 4 and 5 in Sect. 5.1 and 5.2). It is important to  stress that,  in $1+1$ dimensions, integrable Hamiltonians are parametrized by $\frac{n(n-1)}{2}$ arbitrary functions of two variables. On the contrary, in $2+1$ dimensions, the moduli spaces of integrable Hamiltonians are finite-dimensional. Furthermore, the results Sect. 6 (Theorems 6 and 7) make it tempting to conjecture that there exits no non-trivial integrable Hamiltonian systems of hydrodynamic type in $3+1$ dimensions.

The analysis of the integrability conditions is considerably simplified after a transformation of  a given
Hamiltonian system into the so-called Godunov, or symmetric, form. This construction is briefly reviewed in Sect. 3.

The necessary information on hydrodynamic reductions and dispersionless Lax pairs is summarized in Sect. 2.

\section{Hydrodynamic reductions and dispersionless Lax pairs}

Applied to a $(2+1)$-dimensional system (\ref{2+1}), the method of hydrodynamic reductions  consists of seeking multi-phase solutions in the form
$$
{\bf u} (x, y, t)={\bf u}(R^1(x, y, t), ..., R^n(x, y, t))
$$
where the `phases' $R^i(x, y, t)$ are required to satisfy a pair of
 $(1+1)$-dimensional systems of hydrodynamic type,
$$
R_{t}^{i}=\nu ^{i}({\bf R})\ R_{y}^{i}, \ \ \ \ \  R_{x}^{i}=\mu^{i}({\bf R})\ R_{y}^{i}.
$$
Solutions of this form,  known as `non-linear interactions of $n$ planar simple waves'  \cite{Sidorov, Bu, P}, have been extensively discussed in gas dynamics;  later, they reappeared in the context of the dispersionless KP hierarchy, see \cite{GibTsa96, GibTsa99} and references therein. Technically, one `decouples' a $(2+1)$-dimensional system (\ref{2+1}) into a pair of commuting $n$-component $(1+1)$-dimensional systems.
Substituting the ansatz ${\bf u}(R^1, ..., R^n)$ into (\ref{2+1})  one obtains
\begin{equation}
(\nu^i I_n+\mu^iA+B)\ \partial_i{\bf u}=0, ~~~~~ i=1, ..., n,
\label{reduction}
\end{equation}
$\partial_i=\partial/\partial R^i$, implying that both characteristic speeds $\nu^i$ and $\mu^i$ satisfy the dispersion relation
\begin{equation}
{\rm det} (\nu I_n+\mu A+B)=0,
\label{disp}
\end{equation}
which defines an algebraic curve of degree $n$ on the $(\nu, \mu)$-plane.
Moreover,  $\nu^i$ and $\mu^i$ have to satisfy the commutativity conditions
\begin{equation}
\frac{\partial_j\nu^i}{\nu^j-\nu^i}=\frac{\partial_j\mu^i}{\mu^j-\mu^i},
\label{comm}
\end{equation}
$i\ne j$, see \cite{Tsarev}. In was observed in \cite{Fer1} that the requirement of the existence of `sufficiently many' hydrodynamic reductions imposes strong restrictions on the system (\ref{2+1}), and provides an efficient classification criterion. To be precise, we will call a system (\ref{2+1}) integrable if, for any $n$,  it possesses infinitely many $n$-component hydrodynamic reductions parametrized by $n$ arbitrary functions of a single variable.
Thus,  integrable systems are required to possess an infinity of $n$-phase  solutions which can be viewed as natural dispersionless analogs of algebro-geometric solutions of soliton equations.

We recall that a system (\ref{2+1}) is said to possess a dispersionless Lax pair
\begin{equation}
\label{Lax2} \psi_{t} = f\left({\bf u},\psi_{y} \right), \qquad
\psi_{x} = g\left({\bf u},\psi_{y} \right),
\end{equation}
 if it can be recovered from the consistency condition $\psi_{xt}=\psi_{tx}$ (we point out that the dependence of $f$ and $g$ on $\psi_y$ is generally non-linear). Lax pairs of this type first appeared in the construction of the universal Whitham hierarchy, see \cite{Kr} and references therein.
 It was observed in \cite{Zakharov} that such non-linear Lax pairs  arise from the usual `solitonic'  Lax pairs  in the dispersionless limit, and  the cases of  polynomial/rational dependence of $f$ and $g$ on $\psi_y$ were investigated. In particular, a Hamiltonian formulation of such systems was uncovered,  requiring a non-local Hamiltonian density.
  It was demonstrated in \cite{Fer1, Fer3} that, for a number of particularly interesting classes of systems, the existence of a dispersionless Lax pair is {\it equivalent} to the existence  of hydrodynamic reductions and, thus, to the integrability.

Setting $\psi_y=p$ and calculating the consistency condition  $\psi_{xt}=\psi_{tx}$ by virtue of (\ref{2+1}), one arrives at the following relations for $f({\bf u}, p)$ and $g({\bf u}, p)$:
\begin{equation}
{\rm grad} f+{\rm grad}  g \, A=0, ~~~~ {\rm grad}\,  g
\left[f_p I_n+g_pA+B\right]=0; \label{Lax1}
\end{equation}
here ${\rm grad}$ is the gradient with respect to $\bf{u}$.  In particular, this shows that $f_p$ and $g_p$ satisfy the dispersion relation (\ref{disp}), and the vector ${\rm grad}\,  g$ belongs to the left characteristic cone of the system (\ref{2+1}). Thus, as $p$ varies, the equations $\nu=f_p, \ \mu=g_p$ parametrize the dispersion curve (\ref{disp}), while ${\rm grad}\,  g$ parametrizes the left characteristic cone.

Throughout this paper we assume that the dispersion relation (\ref{disp}) defines an {\it irreducible algebraic curve}.  This condition is satisfied for most  examples discussed  in the literature so far.

\section{Transformation of a Hamiltonian system into Godunov's form}

Recall that a system of hydrodynamic type (\ref{2+1})
is said to be symmetrizable, or reducible to Godunov's form \cite{Godunov}, if it possesses a conservative
representation of the form
$$
(\partial_{u^i} p)_t+(\partial_{u^i}q)_x+(\partial_{u^i}r)_y=0;
$$
here the potentials $p, q$ and $ r$ are certain functions of ${\bf u}$. Any such
system possesses an extra conservation law $L(p)_t+L(q)_x+L(r)_y=0$ where $L$ denotes  Legendre's transform.
Equations in Godunov's form play  important role in the general theory of multi-dimensional hyperbolic
conservation laws \cite{Dafermos}.

Given a Hamiltonian system (\ref{Ham})
we  perform the Legendre transform, $H =L(h)= u^ih_i-h,  \ H_i=u^i, \ u_i=h_i,$ to obtain a system  in Godunov's form,
$$
(H_i)_t+(u_i)_x+\lambda^i (u_i)_y=0,
$$
which corresponds to the choice $p=H, \ q=\sum u_i^2/2, \ r=\sum \lambda^iu_i^2/2.$ We assume that the Legendre
transform is well-defined, that is, all partial derivatives $h_i$ are functionally independent. This condition  is
equivalent to the requirement that the Hessian matrix of $h$ is non-degenerate, which is automatically satisfied
under the assumption of  the irreducibility of the  dispersion relation. It turns out that the integrability
conditions take much simpler form when represented in terms of the Legendre transform $H=L(h)$, rather then the Hamiltonian
density $h$ itself. Thus, in what follows we will work with systems represented in Godunov's form (to make
the equations look formally `evolutionary'  we will relabel the independent variables as $x, y, t\to T, X, Y$).
This results in
\begin{equation}
(u_i)_T+\lambda^i (u_i)_X+(H_i)_Y=0; \label{G}
\end{equation}
Systems of this type can be viewed as describing $n$ linear waves
(traveling with constant speeds $\lambda^i$ in the $X,T$-plane)
which are non-linearly coupled in the $Y$-direction.

\section{Integrable Hamiltonians in $2+1$ dimensions: two-component case}

In this section we classify two-component Hamiltonian systems (\ref{2comp}). The corresponding Legendre transform is
\begin{equation}
v_T+(H_v)_Y=0, ~~~ w_X+(H_w)_Y=0;
\label{vw}
\end{equation}
 here $v=u_1, w=u_2$.
  We point out that this case was addressed previously in \cite{Fer1},  although the classification was only sketched.
  Here  we  provide a complete list of integrable potentials $H(v, w)$, and calculate the corresponding dispersionless
  Lax pairs. For systems (\ref{vw})  the integrability conditions   constitute an over-determined system of fourth
  order PDEs for the potential $H(v, w)$:
\begin{equation}
\begin{array}{c}
H_{vw}H_{vvvv}=2H_{vvv}H_{vvw}, \\
\ \\
H_{vw}H_{vvvw}=2H_{vvv}H_{vww}, \\
\ \\
H_{vw}H_{vvww}=H_{vvw}H_{vww}+H_{vvv}H_{www}, \\
\ \\
H_{vw}H_{vwww}=2H_{vvw}H_{www}, \\
\ \\
H_{vw}H_{wwww}=2H_{vww}H_{www}.
\end{array}
\label{H}
\end{equation}
The system (\ref{H}) is in involution, and its solution space is $10$-dimensional  \cite{Fer1}.  We point out that
the transformations
$$
v\to av+b, ~~~ w\to cw+d, ~~~ H\to \alpha H +\beta v^2+\gamma w^2 + \mu v+\nu w+\delta
$$
generate a $10$-dimensional  group of Lie-point symmetries of the system (\ref{H}). These transformations correspond
to obvious linear changes of the independent variables $X, Y, T$ in the equations (\ref{vw}). One can show that
the action of the symmetry group on the moduli space of solutions of the system (\ref{H}) possesses an open orbit.
The classification of integrable potentials $H(v, w)$ will be performed up to this equivalence. Moreover, we will
not be interested  in the potentials which are either quadratic in $v, w$ and generate linear systems (\ref{vw}),
or   separable potentials of the form $f(v)+g(w)$  giving rise to reducible systems.
Our main result is the following complete list of integrable potentials:

\begin{theorem}
The `generic' solution of the system (\ref{H}) is given by the formula
\begin{equation}
H(v, w)=Z(v+w)+\epsilon Z(v+\epsilon w)+\epsilon^2 Z(v+\epsilon^2
w);
 \label{gen}
\end{equation}
here $\epsilon=e^{2\pi i/3}$ and $Z''(s)=\zeta(s)$ where $\zeta $ is the
Weierstrass zeta-function: $\zeta'=-\wp, \ (\wp')^2=4\wp^3-g_3$. Degenerations of this solution  correspond to
\begin{equation}
H(v, w)=\frac{1}{2}v^2\zeta (w),  \label{gen1}
\end{equation}
\begin{equation}
H(v, w)=(v+w)\ln (v+w), \label{gen11}
\end{equation}
as well as the following  polynomial potentials:
\begin{equation}
H(v, w)=v^2w^2,
\label{gen2}
\end{equation}
\begin{equation}
H(v, w)=vw^2+\frac{\alpha}{5}w^5, ~~~ \alpha=const,
\label{gen3}
\end{equation}
and
\begin{equation}
H(v, w)=vw+\frac{1}{6}w^3.
\label{gen4}
\end{equation}
\end{theorem}
\noindent {\bf Remark.} The `elliptic'  examples (\ref{gen}) and  (\ref{gen1}) possess a specialization  $g_3=0$:
$\wp(w)\to 1/w^2, ~ \zeta (w) \to 1/w,  ~ \sigma (w) \to w,$ etc. This results in the potentials
\begin{equation}
H(v, w)=(v+w) \log(v+w)+\epsilon (v+\epsilon w) \log(v+\epsilon
w)+\epsilon^2 (v+\epsilon^2 w) \log(v+\epsilon^2 w)
\label{deg}
\end{equation}
and
$$
H(v, w)=\frac{v^2}{2w},
$$
respectively. Dispersionless Lax pairs for the equations
(\ref{vw}) corresponding to the potentials
(\ref{gen})-(\ref{gen4}) are calculated in Sect. 4.1.

\medskip

\centerline{\bf Proof of Theorem 1:}

\medskip

The system (\ref{H}) can be solved as follows. The first two  equations  imply that $H_{vvv}/H_{vw}^2={\rm const}$.
Similarly, the last two equations imply $H_{www}/H_{vw}^2={\rm
const}$. Setting $H_{vw}=e$ one can parametrise the third order derivatives of $H$ in the form
\begin{equation}
H_{vvv}=\frac{1}{2}me^2, ~~~ H_{vvw}=e_v, ~~~ H_{vww}=e_w, ~~~
H_{www}=\frac{1}{2}ne^2, \label{H1}
\end{equation}
here $m, n $ are arbitrary constants. The compatibility conditions
for these equations, plus the equation $(\ref{H})_3$,  result
in the following  overdetermined system for $e$:
\begin{equation}
(\ln e)_{vw}=\frac{mn}{4}e^2, ~~~ e_{vv}=mee_{w}, ~~~ e_{ww}=nee_v.
\label{e}
\end{equation}

\noindent The general solution of the first (Liouville) equation  has the form
\begin{equation}
e^2=\frac{4}{mn}\frac{p'(v)q'(w)}{(p(v)+q(w))^2}, \label{L}
\end{equation}
one has to consider separately the case $e=const$ (up to the equivalence transformations,  this results in the potential
(\ref{gen4})), as well as the case when $e$ depends on one variable only, say, on $w$ (this leads to the potential
(\ref{gen3})). Let us
assume that both constants $m$ and $n$ are nonzero (the cases when either  of them vanishes will be discussed later).
By scaling $v$ and $w$  one can assume $m=n=1$.
Setting
\begin{equation}
(p')^{3}= P^2(p), ~~~ (q')^{3}= Q^2(q),
\label{pq}
\end{equation}
(here $P(p)$ and $Q(q)$ are functions to be determined),  one obtains
from  the last two equations (\ref{e}) the following
functional-differential equations for $P$ and $Q$:
$$
P''(p+q)^2-4P'(p+q)+6P=2Q'(p+q)-6Q,  ~~~
Q''(p+q)^2-4Q'(p+q)+6Q=2P'(p+q)-6P;
$$
these equations  imply that
both $P$ and $Q$ are cubic polynomials in $p$ and $q$,
$$
P=ap^3+bp^2+cp+d, ~~~ Q=aq^3-bq^2+cq-d,
$$
where $a, b, c, d$ are arbitrary constants.  Notice that the right hand side of (\ref{L}) possesses the
following $SL(2, R)$-invariance,
$$
p\to \frac{\alpha p+\beta}{\gamma p+\delta}, ~~~ q\to -\frac{\alpha p-\beta}{\gamma p-\delta},
$$
which can be used to bring the polynomials $P(p)$ and $Q(q)$ to canonical forms. There are three cases to consider.

\noindent {\bf Three distinct roots:}  in this case one can reduce both $P(p)$ and $Q(q)$ to quadratics, so that
the ODEs (\ref{pq}) assume the form
$$
(p')^{3}=\frac{27}{2}(p^2+g_3)^2 ~~~ {\rm and} ~~~  (q')^{3}=\frac{27}{2}(q^2+g_3)^2,
$$
respectively. Thus, $p=\wp'(v), \ q=\wp'(w)$ where $\wp$ is the Weierstrass $\wp$-function: $(\wp')^2=4\wp^3-g_3$ (we point
out that the value of $g_3$ is not really essential, and can be normalized to $\pm 1$). Setting
$$
H_{vw}=e=- \frac{12\wp(v)\wp(w)}{\wp'(v)+\wp'(w)}
$$
and  integrating (\ref{H1}) with respect to $v$ and $w$ we obtain
\begin{equation}
H_{vv}=-6\zeta(w)- \frac{12\wp^2(w)}{\wp'(v)+\wp'(w)}, ~~ H_{vw}=
-\frac{12\wp(v)\wp(w)}{\wp'(v)+\wp'(w)}, ~~ H_{ww}=-6\zeta(v)-
\frac{12\wp^2(v)}{\wp'(v)+\wp'(w)}, \label{H11}
\end{equation}
here  the zeta-function is defined as $\zeta'=-\wp$. Since the $\wp$-function on the elliptic
curve $y^2=4 x^3-g_3$ satisfies the  automorphic property
$\wp(\epsilon z)=\epsilon \wp(z), \ \epsilon^3=1$, one can  rewrite (\ref{H11})
 in the following equivalent form:
$$
H_{vv}=-2 \Big(\zeta(v+w)+\epsilon \zeta(v+\epsilon
w)+\epsilon^2 \zeta(v+\epsilon^2 w)\Big),
$$
$$
H_{vw}=-2 \Big(\zeta(v+w)+\epsilon^2 \zeta(v+\epsilon
w)+\epsilon \zeta(v+\epsilon^2 w)\Big),
$$
$$
H_{ww}=-2 \Big(\zeta(v+w)+\zeta(v+\epsilon w)+
\zeta(v+\epsilon^2 w)\Big).
$$
Up to a constant multiple, these formulae give rise to (\ref{gen}).

\noindent {\bf Double root:}  in this case  both $P(p)$ and $Q(q)$ can be reduced to $p$ and $q$, so that the
ODEs (\ref{pq}) take the form $(p')^{3}=27p^2$ and $(q')^{3}= 27q^2$, respectively. This leads to $p=v^3, \ q=w^3$,
and a straightforward integration of  (\ref{H1}) gives
$$
H_{vv}=-\frac{6w^2}{v^3+w^3}, ~~ H_{vw}=\frac{6vw}{v^3+w^3}, ~~ H_{ww}=-\frac{6v^2}{v^3+w^3};
$$
notice that these formulae can be obtained as a degeneration of (\ref{H11}) corresponding to $g_3=0$.
Up to a constant multiple, this leads to the potential (\ref{deg}).

\noindent {\bf Triple root:} in this case  both $P(p)$ and $Q(q)$ can be reduced to constants, so that the ODEs
(\ref{pq}) take the form $(p')^{3}=1$ and $(q')^{3}= 1$, respectively. This leads to $e=2/(v+w)$, which, up to a
constant multiple, results in the potential (\ref{gen11}).

 If $m=0, \ n\ne 0$ (without any loss of generality we will again set $n=1$), equations (\ref{e})
 can be solved in the form $e=6v\wp(w)$ where $\wp$ is the Weierstrass $\wp$-function: $(\wp')^2=4\wp^3-g_3$.
 The corresponding potential $H$ is given by
$
H=-3 v^2 \zeta(w).
$
Up to a multiple, this is the case (\ref{gen1}).

In the simplest case $m=n=0$ equations (\ref{e}) imply
$$
e=(\alpha v +\beta)( \gamma w +\delta),
$$
and the elementary integration  of  equations (\ref{H1}) results in
$$
H(v, w)=(\frac{1}{2} \alpha v^2+\beta v)(\frac{1}{2} \gamma w^2+\delta w);
$$
here $\alpha, \beta, \gamma, \delta$ are arbitrary constants. Using
the equivalence transformations  one can reduce $H$
to either $H=v^2w^2$ (both $\alpha$ and $\gamma$ are nonzero) or
$H=vw^2$ ($\alpha =0$). These are the polynomial cases (\ref{gen2}) and a subcase of (\ref{gen3}), respectively. This finishes the proof of Theorem 1.

\subsection{Dispersionless Lax pairs}

In this section we calculate dispersionless Lax pairs for  systems (\ref{vw}) corresponding to the potentials (\ref{gen})-(\ref{gen4}) of Theorem 1.  We point out that, in spite of the deceptive simplicity of some of these potentials, the corresponding  Lax pairs are quite non-trivial.

\medskip

\noindent {\bf Potential (\ref{gen4}):}
The corresponding system (\ref{vw}) takes the form
\begin{equation}
v_T+w_Y=0, ~~~ w_X+ww_Y+v_Y=0;
\label{s3}
\end{equation}
it arises in the genus zero case of the universal Whitham hierarchy \cite{Kr, Manas}.  This system possesses the Lax pair
$$
\psi_T=\frac{1}{2}\ln (\psi_Y+w/2), ~~~ \psi_X=\psi_Y^2+v/2.
$$
A simple calculation shows that the Legendre transform of the potential $H(v, w)=vw+\frac{1}{6}w^3$, defined by the
formulae
$$
u^1=H_v, ~~~ u^2=H_w, ~~~ h(u^1, u^2)=vH_v+wH_w-h,
$$
is also polynomial:
$$
h(u^1, u^2)=u^1u^2-\frac{1}{6}(u^1)^3.
$$
We point out that all other examples of integrable potentials $H(v, w)$ produce non-polynomial Hamiltonian
densities $h(u^1, u^2)$.

\medskip

\noindent {\bf Potential (\ref{gen3}):}
The corresponding system (\ref{vw}) takes the form
\begin{equation}
v_T+(w^2)_Y=0, ~~~ w_X+2(vw)_Y+\alpha (w^4)_Y=0.
\label{s2}
\end{equation}
For $\alpha=0$ it possesses the Lax pair
$$
\psi_T=-\frac{w^2}{2\psi_Y^2}, ~~~ \psi_X=\psi_Y^4-2v\psi_Y.
$$
Setting $v=u_Y, \ w^2=-u_T$ one can rewrite (\ref{s2}) (when $\alpha=0$) as a single second order PDE
$$
u_{XT}+2u_Yu_{TY}+4u_Tu_{YY}=0.
$$
Up to a  rescaling  $X\to -2X$ this equation is a particular case of the generalized dispersionless Harry Dym
equation \cite{Bla, Pavlov}.
For $\alpha \ne 0$ the Lax pair modifies to
$$
\psi_T=f\left(\frac{w}{\psi_Y}\right), ~~~ \psi_X=\psi_Y^4-2v\psi_Y,
$$
where the function $f(s)$ satisfies the equation $f'(s)=-s/(\alpha
s^3+1)$ (for $\alpha=0$ one recovers the previous formula).  The
first equation of this Lax pair appeared in \cite{Pavlov} as a
generating function of conservation laws for the Kupershmidt
hydrodynamic chain. Without any loss of generality one can set
$\alpha=-1$, which gives
$$
f(s)=\frac{1}{3}\left(\ln (s-1)+\epsilon^2\ln(s-\epsilon)+\epsilon\ln(s-\epsilon^2)\right), ~~~ \epsilon^3=1.
$$

\medskip

\noindent {\bf Potential (\ref{gen2}):}
The corresponding system (\ref{vw})
takes the form
\begin{equation}
v_T+2(vw^2)_Y=0, ~~~ w_X+2(v^2w)_Y=0.
\label{s1}
\end{equation}
It possesses the Lax pair
$$
\psi_T=w^2 a(\psi_Y), ~~~ \psi_X=-v^2 b(\psi_Y)
$$
where the dependence of $a$ and $b$ on $\psi_y\equiv \xi$
is governed by
the ODEs
$$
a'=-4\frac{a}{b}-2, ~~~ b'=4\frac{b}{a}+2.
$$
To solve these equations we proceed as follows. Expressing $b$
from the first equation, $b=-4a/(a'+2)$, and substituting into the
second one arrives at a second order ODE $2aa''-3(a')^2+12=0$. It
can be integrated once, $(a')^2=4ca^3+4$, where c is a constant of
integration. Without any loss of generality we will set $c=1$.
Thus, $a$ is the Weierstrass $\wp$-function: $a=\wp(\xi, 0,
-4)=\wp(\xi)$. The corresponding $b$ is given by
$b=-4\wp/(\wp'+2)$. Notice that this expression for $b$ equals
$\wp(\xi+c)$ where $c$ is the zero of $\wp$-function such that
$\wp(c)=0, \ \wp'(c)=2$ (use the addition theorem to calculate
$\wp(\xi+c)$). Ultimately, we obtain the Lax pair
$$
\psi_T=w^2 \wp(\psi_Y), ~~~ \psi_X=-v^2 \wp(\psi_Y+c).
$$
Setting $V=v^2, \ W=w^2$ one can rewrite (\ref{s1}) in the form where the non-linearity is quadratic:
$$
V_t+2WV_Y+4VW_Y=0, ~~~ W_x+2VW_Y+4WV_Y=0.
$$

\medskip

\noindent {\bf Potential (\ref{gen11}):}
The corresponding system (\ref{vw})
takes the form
$$
v_T+\frac{v_Y+w_Y}{v+w}=0, ~~~ w_X+\frac{v_Y+w_Y}{v+w}=0.
$$
It possesses the Lax pair
$$
\psi_T=-\ln(w+\psi_Y), ~~~ \psi_X=\ln(v-\psi_Y).
$$
This system also arises in the genus zero case of the universal Whitham hierarchy \cite{Kr, Manas}; its
dispersionful analogue was constructed in \cite{B}.

\medskip

\noindent {\bf Potential (\ref{gen1}):} The corresponding system (\ref{vw})
takes the form
$$
v_T+\zeta(w)v_Y-v\wp(w)w_Y=0, ~~~
w_X-\wp(w)vv_Y-\frac{1}{2}v^2\wp'(w) w_Y=0.
$$
One can show that it possesses the Lax pair
$$
\psi_T=-f(w, \psi_Y), ~~~ \psi_X=-\frac{1}{2}v^2b(\psi_Y)
$$
where, setting $\psi_Y\equiv \xi$, the function  $f(w, \xi)$ has to satisfy the equations
$$
f_w=\frac{2b(\xi)\wp(w)}{b'(\xi)+\wp'(w)}, ~~~ f_{\xi}=\zeta(w)+ \frac{2\wp^2(w)}{b'(\xi)+\wp'(w)}.
$$
We point out that the consistency condition $f_{\xi w}=f_{w\xi}$ implies a second order ODE
$2bb''-3(b')^2-3g_3=0$ which, upon integration, gives
$$
(b'(\xi))^2=4b^3(\xi)-g_3,
$$
(the constant of integration is not essential). Thus, one can set $b=\wp(\xi)$ so that the equations for $f$
take the form
$$
f_w=\frac{2\wp(\xi)\wp(w)}{\wp'(\xi)+\wp'(w)}, ~~~ f_{\xi}=\zeta(w)+ \frac{2\wp^2(w)}{\wp'(\xi)+\wp'(w)},
$$
compare with (\ref{H11})! Thus,
$$
f(w, \xi)=\frac{1}{3}\ln \sigma(\xi+w)+\frac{\epsilon}{3} \ln \sigma(\xi+\epsilon w)+\frac{\epsilon^2}{3}
\ln \sigma(\xi+\epsilon^2w),
$$
where $\sigma$ is the Weierstrass sigma-function: $\sigma'/\sigma=\zeta$. Ultimately,  the Lax pair takes the form
$$
\psi_T=\frac{1}{3}\ln \sigma(\psi_Y+w)+\frac{\epsilon}{3} \ln \sigma(\psi_Y+\epsilon w)+\frac{\epsilon^2}{3}
\ln \sigma(\psi_y+\epsilon^2w), ~~~ \psi_X=-\frac{1}{2}v^2\wp(\psi_Y).
$$

\medskip

\noindent {\bf Potential (\ref{gen}):} the equations corresponding to $H/3$ take the form
$$
\begin{array}{c}
v_T+\left(\zeta(w)+ \frac{2\wp^2(w)}{\wp'(v)+\wp'(w)}\right) v_Y+\frac{2\wp(v)\wp(w)}{\wp'(v)+\wp'(w)} w_Y=0, \\
\ \\
w_X+\frac{2\wp(v)\wp(w)}{\wp'(v)+\wp'(w)} v_Y+\left(\zeta(v)+ \frac{2\wp^2(v)}{\wp'(v)+\wp'(w)}\right) w_Y=0.
\end{array}
$$
One can show that the corresponding Lax pair is given by the equations
$$
\psi_T=f(w, \psi_Y), ~~~ \psi_X=g(v, \psi_Y)
$$
where, setting $\psi_Y=\xi$, the first order partial derivatives of $f$ and $g$ are given by
$$
f_w=-\frac{2\wp(\xi)\wp(w)}{\wp'(\xi)+\wp'(w)}, ~~~ f_{\xi}=-\zeta(w)- \frac{2\wp^2(w)}{\wp'(\xi)+\wp'(w)}
$$
and
$$
g_v=-\frac{2\wp(\xi)\wp(v)}{\wp'(\xi)-\wp'(v)}, ~~~ g_{\xi}=-\zeta(v)+ \frac{2\wp^2(v)}{\wp'(\xi)-\wp'(v)},
$$
respectively. Explicitly, one has
$$
f(w, \xi)=-\frac{1}{3}\ln \sigma(\xi+w)-\frac{\epsilon}{3} \ln \sigma(\xi+\epsilon w)-\frac{\epsilon^2}{3}
\ln \sigma(\xi+\epsilon^2w),
$$
$$
g(v, \xi)=\frac{1}{3}\ln \sigma(\xi-v)+\frac{\epsilon}{3} \ln \sigma(\xi-\epsilon v)+\frac{\epsilon^2}{3}
\ln \sigma(\xi-\epsilon^2v).
$$
Notice that the expression for $f(w, \xi)$ coincides with the one from the previous case. This means that the corresponding Hamiltonian systems commute with each other --- the fact which is, in a sense, unexpected.

\medskip

\noindent {\bf Potential (\ref{deg}):} this is the $g_3=0$ degeneration of the potential (\ref{gen}). The
system corresponding to $H/3$ takes the form
$$
v_T+\frac{w^2}{v^3+w^3}v_Y-\frac{vw}{v^3+w^3}w_Y=0, ~~~
w_X-\frac{vw}{v^3+w^3}v_Y+\frac{v^2}{v^3+w^3}w_Y=0;
$$
it possesses the Lax pair
$$
\psi_T=f(w/\psi_Y), ~~~ \psi_X=g(v/\psi_Y)
$$
where the dependence of $f$ and $g$ on their arguments is specified by
$f'(s)=s/(s^3-1), \ g'(s)=s/(s^3+1)$. Explicitly, one has
$$
f(s)=\frac{1}{3}\left(\ln (s-1)+\epsilon^2\ln(s-\epsilon)+\epsilon\ln(s-\epsilon^2)\right),
$$
$$
g(s)=-\frac{1}{3}\left(\ln (s+1)+\epsilon^2\ln(s+\epsilon)+\epsilon\ln(s+\epsilon^2)\right).
$$

\section{Integrable Hamiltonians in $2+1$ dimensions: three-component case}

In this section we classify three-component integrable equations of the form (\ref{G}),
\begin{equation}
\label{Hessian}
\left(\begin{array}{c}
u_1 \\
u_2\\
u_3
\end{array}
\right)_T+
\left(\begin{array}{ccc}
\lambda^{1} & 0 & 0\\
0 & \lambda^{2} & 0\\
0&0&\lambda^{3}
\end{array}
\right)
\left(\begin{array}{c}
u_1 \\
u_2\\
u_3
\end{array}
\right)_X+
\left(\begin{array}{ccc}
H_{11} & H_{12} & H_{13}\\
H_{12} & H_{22} & H_{23}\\
H_{13}& H_{23}& H_{33}
\end{array}
\right)
\left(\begin{array}{c}
u_1 \\
u_2\\
u_3
\end{array}
\right)_Y=0,
\end{equation}
assuming that the constants $\lambda^i$ are pairwise distinct. As
mentioned in the introduction, the integrability of the
system~(\ref{Hessian}) implies the vanishing of the Haantjes
tensor for any matrix of the two-parameter family  $(kA +
I_{3})^{-1} (lB + I_{3})$. Here $A=diag(\lambda^i)$ and
$B=(H_{ij})$. To formulate the integrability conditions in a
compact form we introduce the following  notation:
\begin{equation*}R_{1} = \frac{H_{12}H_{13}}{H_{23}} \left(\lambda^{2} - \lambda^{3} \right), \qquad
R_{2} = \frac{H_{12}H_{23}}{H_{13}} \left(\lambda^{3} -
\lambda^{1} \right), \qquad R_{3} = \frac{H_{13}H_{23}}{H_{12}}
\left(\lambda^{1} - \lambda^{2} \right);
\end{equation*}
 we will see below that  all mixed partial derivatives $H_{ij}$ must be non-zero, otherwise the system is either linear, or reducible. Moreover, we will need the quantities
\begin{equation*}
I = \Delta^{2} - 4 (\lambda^{2} - \lambda^{3}) (\lambda^{3} -
\lambda^{1}) H_{12}^{2} - 4 (\lambda^{3}-\lambda^{1})
(\lambda^{1}-\lambda^{2}) H_{23}^{2} - 4 (\lambda^{1}-\lambda^{2})
(\lambda^{2}-\lambda^{3}) H_{13}^{2}
\end{equation*}
and
\begin{equation*}
J = (\lambda^{2} - \lambda^{3}) H_{12}^{2} H_{13}^{2}
+(\lambda^{3} - \lambda^{1}) H_{23}^{2} H_{12}^{2} +
(\lambda^{1}-\lambda^{2}) H_{13}^{2} H_{23}^{2} - H_{12} H_{23}
H_{13} \Delta
\end{equation*}
where
\begin{equation*}
\label{delta} \Delta = (\lambda^{2} - \lambda^{3}) H_{11} +
(\lambda^{3} - \lambda^{1}) H_{22} + (\lambda^{1} - \lambda^{2})
H_{33}.
\end{equation*}
Our first result is the following

\begin{theorem}
\label{theoDiag}The system (\ref{Hessian}) with an irreducible
dispersion curve is diagonalizable if and only if the potential
$H$ satisfies the relations
\end{theorem}
\begin{equation}
\label{H123}
\begin{array}{c}
J=0, ~~~ H_{123} = 0, \\
\ \\
\frac{\partial}{\partial u_{1}} \left((\lambda^{3} - \lambda^{2})
H_{11} + R_{2} + R_{3} \right) = 0, \\
\ \\
\frac{\partial}{\partial u_{2}} \left((\lambda^{1} - \lambda^{3})
H_{22} + R_{1} + R_{3} \right) = 0, \\
\ \\
\frac{\partial}{\partial u_{3}} \left((\lambda^{2} - \lambda^{1})
H_{33} + R_{1} + R_{2} \right) = 0.
\end{array}
\end{equation}
Notice that, in contrast to the two-component situation (\ref{H}), these relations are third order in the
derivatives of $H$. We will demonstrate below that the necessary conditions (\ref{H123}) are, in fact, sufficient for the integrability, and imply the existence
of dispersionless Lax pairs and an infinity of hydrodynamic reductions.

\noindent {\bf Remark.} The condition $J=0$, which is equivalent to $R_1+R_2+R_3=\Delta$, has a simple geometric interpretation as the condition of reducibility of the left characteristic cone of the system (\ref{Hessian}) (see Sect. 2 for  definitions). Indeed, the left characteristic  cone consists of all vectors
${\bf g}=(g_1, g_2, g_3)$ which satisfy the relation
\begin{equation}
{\bf g}(\nu I_3+\mu A+B)=0.
\label{dual}
\end{equation}
Excluding $\nu$ and $\mu$, one obtains a single algebraic relation among $g_1, g_2, g_3$,
\begin{gather}
\begin{aligned}
\label{cone} \left(H_{13} (g_{1})^{2} g_{2} + H_{23} g_{1}
(g_{2})^{2}  +
 H_{33} g_{1} g_{2} g_{3}  \right) \left (\lambda^{1} - \lambda^{2}\right) &+& \\
\left(H_{21} (g_{2})^{2} g_{3} + H_{13} g_{2} (g_{3})^{2}  + H_{11} g_{1} g_{2} g_{3}  \right) \left (\lambda^{2} - \lambda^{3}\right) &+& \\
\left(H_{23} (g_{3})^{2} g_{1} + H_{12} g_{3} (g_{1})^{2}  +
H_{22} g_{1} g_{2} g_{3}  \right) \left (\lambda^{3} -
\lambda^{1}\right) &=& 0,
\end{aligned}
\end{gather}
which is the equation of the left characteristic cone. The condition $J=0$ is
equivalent to its degeneration into  a line and a conic:
\begin{gather}
\begin{aligned}
&\left[H_{12} H_{13} g_{1} + H_{12} H_{23} g_{2} + H_{13} H_{23} g_{3}\right]\\
&\left[H_{13} H_{23} (\lambda^{1} - \lambda^{2}) g_{1} g_{2}
+H_{12} H_{23} (\lambda^{3} - \lambda^{1}) g_{1} g_{3} + H_{12}
H_{13} (\lambda^{2} - \lambda^{3}) g_{2} g_{3}\right]=0.
\label{decomp}
\end{aligned}
\end{gather}
We point out that, by virtue of (\ref{dual}), the left
characteristic cone and the dispersion curve are birationally
equivalent. This implies that the dispersion curve is necessarily
rational, although not  reducible (the linear factor of the left
characteristic cone corresponds to a singular point on the
dispersion curve --- see Sect. 5.2 for explicit formulae).

\medskip

\centerline{\bf Proof of Theorem~\ref{theoDiag}:}

\medskip

To simplify the calculation of the Haantjes tensor we multiply the
matrix  $(kA + I_{3})^{-1} (lB + I_{3})$ by
$(k\lambda^1+1)(k\lambda^2+1)(k\lambda^3+1)$. This results in the
matrix $\tilde{A} (l B + I_3)$ where $\tilde{A} = diag
[(k\lambda^2+1)(k\lambda^3+1), \ (k\lambda^1+1)(k\lambda^3+1), \
(k\lambda^1+1)(k\lambda^2+1)]$. Since the multiplication by a
scalar does not effect the vanishing of the Haantjes tensor, we
will work with the matrix $\tilde{A} (lB + I_{3})$
 which has an advantage of being polynomial in $k$ and $l$.
Using computer algebra we calculate  components of the Haantjes
tensor ${\cal H}$ (which are certain polynomials in $k$ and $l$) and
set them equal to zero. First of all, one can verify that all
components of the form ${\cal H}_{ij}^{i} $ vanish identically, so
that the only nonzero components are ${\cal H}_{jk}^{i}, \ i\ne j\ne
k$. In the following we will focus on the analysis of the component
${\cal H}_{12}^{3}$: it turns out  the vanishing of  ${\cal
H}_{12}^{3}$ alone implies the vanishing of the full Haantjes tensor. Let
us compute coefficients at different powers of the parameter $l$ and
set them equal to zero. At the order $l^{0}$, all terms in ${\cal
H}_{12}^{3}$
  vanish identically since $\tilde{A}$ is a
constant diagonal matrix. The coefficient at $l^1$ is a polynomial
in $k$, however, setting  its coefficients equal to zero we obtain
only one independent relation:
\begin{equation*}
H_{123} = 0.
\end{equation*}
Similarly,  two extra  relations come from the analysis of
$l^{2}$-terms, three relations from $l^{3}$-terms, and four
relations from $l^{4}$-terms. Ultimately, we end up with a set of
$9$ linear homogeneous equations for the $9$ third order derivatives $H_{iii}, \  H_{iij}$.

From these $9$ relations it readily follows that if one of the mixed derivatives equals zero,
say, $H_{12}=0$, then either $H_{13} H_{23}=0$ or
$H_{ijk}=0$ for all $i,j,k$. In the first case the system
(\ref{Hessian})  decouples into a pair of  independent $1\times 1$ and $2 \times 2$ subsystems. The second case corresponds to linear systems
with constant coefficients. Therefore, from now on we assume  $H_{ij}\ne 0$
 for any $i\ne j$.

The set of $9$ relations so obtained is rather complicated, and the
calculation of the corresponding $9 \times 9$ determinant is
computationally intense. A simpler equivalent set of relations can
be derived  as follows: first, divide ${\cal H}_{12}^{3}$ by
$(\lambda^{1} k + 1) (\lambda^{2} k + 1)^{2} (\lambda^{3} k +
1)^{2}$ (which is a common multiple), then equate to zero the
coefficient of $l^{2}$ at $k = -1/\lambda^{1}, -1/\lambda^{2}$ (the
coefficient at $k = -1/\lambda^{3}$ appears to be a linear
combination of the previous two), the coefficient of $l^{3}$ at $k =
-1/\lambda^{1}, -1/\lambda^{2},-1/\lambda^{3}$ and the coefficient
of $l^{4}$ at $k =0, -1/\lambda^{1}, -1/\lambda^{2},-1/\lambda^{3}$.
As a result, we arrive at a simpler set of $9$ linearly independent
relations that are nothing but  linear combinations of the previous
ones. If the determinant of this system is non-zero, then all
remaining derivatives $H_{iii}$ and $ H_{iij}$ vanish identically. This is the case
of linear systems. Thus,  to obtain non-linear examples, one has to require the vanishing of the
determinant. It is straightforward to
verify that this  determinant factorizes as follows:
\begin{equation*}
J^4 \, \Big(I^{2} - 64 (\lambda^{1} - \lambda^{2}) (\lambda^{2} -
\lambda^{3}) (\lambda^{3} - \lambda^{1}) J \Big)= 0.
\end{equation*}
Thus, there are two cases to consider. If
\begin{equation}
\label{IJ64}
 I^{2} -
64 (\lambda^{1} - \lambda^{2}) (\lambda^{2} - \lambda^{3})
(\lambda^{3} - \lambda^{1}) J = 0,
\end{equation}
then  the dispersion relation of the system~(\ref{Hessian}) is
reducible. To show this we  introduce the quantities
\begin{align*}
\Omega_{1} &= \Delta H_{12} - 2 H_{13} H_{23} (\lambda^{1}
-\lambda^{2}), \\
\Omega_{2} &= \Delta H_{23} - 2 H_{12} H_{13} (\lambda^{2}
-\lambda^{3}), \\
\Omega_{3} &= \Delta^{2} - 4 H_{13}^{2} (\lambda^{1} -
\lambda^{2})
(\lambda^{2} - \lambda^{3}), \\
\Omega_{4} &= H_{12}^{2} (\lambda^{3} -\lambda^{2}) + H_{23}^{2}
(\lambda^{1} -\lambda^{2}),
\end{align*}
which can be verified to satisfy the quadratic identity
\begin{equation}
\label{OmegaId} (\lambda^{2} - \lambda^{3}) \Omega_{1}^{2} +
(\lambda^{2} -\lambda^{1}) \Omega_{2}^{2} + \Omega_{3} \Omega_{4}
=0.
\end{equation}
In terms of these quantities, the equation~(\ref{IJ64}) can be rewritten as follows:
\begin{equation}
\left( \Omega_{3} - 4 (\lambda^{1} - \lambda^{3}) \Omega_{4}
\right)^{2} + 16 (\lambda^{1} -\lambda^{2}) (\lambda^{1} -
\lambda^{3}) \Omega_{2}^{2} = 0,
\label{A1}
\end{equation}
or, equivalently,
\begin{equation}
\left( \Omega_{3} + 4 (\lambda^{1} - \lambda^{3}) \Omega_{4}
\right)^{2} + 16 (\lambda^{1} -\lambda^{3}) (\lambda^{2} -
\lambda^{3}) \Omega_{1}^{2} = 0;
\label{A2}
\end{equation}
one has to use the identity (\ref{OmegaId}) to verify the equivalence of (\ref{A1}) and (\ref{A2}). Let us
assume that $\lambda^{1} < \lambda^{2} <
\lambda^{3}$. Since we are interested in real-valued solutions, the equation (\ref{A1}) implies
\begin{equation}
 \label{OmegaEq1} \Omega_{2} = 0,
\qquad \Omega_{3} = 4 (\lambda^{1} - \lambda^{3}) \Omega_{4};
\end{equation}
(one should use  (\ref{A2}) if $\lambda^{2} < \lambda^{1} <
\lambda^{3}$ ). In this case  the identity (\ref{OmegaId}) takes the form
$$
(\lambda^{2}-\lambda^{3}) \Omega_{1}^{2} + 4
(\lambda^{1}-\lambda^{3}) \Omega_{4}^{2} = 0,
$$
so that
$$
\Omega_{1} =0, \qquad \Omega_{4} = 0.
$$
These conditions lead to potentials of the form
$$
H=u_2(\gamma u_1+\delta u_3)+f(\gamma u_1+\delta u_3);
$$
here the constants $\gamma$ and $\delta$ satisfy the relation
$(\lambda^2-\lambda^1)\delta^2+(\lambda^2-\lambda^3)\gamma^2=0$, and $f$ is an arbitrary function of the indicated argument.
This ansatz, however, implies the reducibility of the dispersion relation as discussed in \cite{Fer2}.
Thus, we are left with the second branch  $J = 0$, in which case the rank of the system drops to $5$, and we end up with the equations (\ref{H123}). This finishes the proof of Theorem 2.

\medskip

The main result of this Section  is a complete list of integrable potentials $H(u_1, u_2, u_3)$ which come from a detailed analysis of the equations (\ref{H123}). The classification will be performed up to the following equivalence transformations, which constitute a group of point symmetries of the relations
(\ref{H123}).

\noindent {\bf Equivalence transformations:}

\noindent transformations of the variables $u_i$: $u_i\to au_i+b_i;$

\noindent transformations of the potential $H$:
$$
H\to \alpha H+ \beta \sum u_i^2/2+ \gamma \sum \lambda^i u_i^2/2+\mu_iu_i+\delta,
$$
the latter corresponding to $Y\to \alpha Y +\beta T +\gamma X$ in the equations (\ref{Hessian}). Moreover,  relations (\ref{H123}) are  invariant under arbitrary permutations of indices.
Finally, we will not be interested  in the potentials which are either quadratic in $u_i$ and generate linear systems (\ref{Hessian}), or  separable potentials, e.g.,  $H=f(u_1)+g(u_2, u_3)$,  giving rise to reducible systems.

\begin{theorem}
\label{solH123} The `generic' solution of the equations
(\ref{H123})  is given by the formula
\begin{equation}
H=-\sum_{j\ne i}\frac{\lambda^i-\lambda^j}{6a^2_ia^2_j}V(a_iu_i, a_ju_j)
\label{31}
\end{equation}
where
\begin{equation}
V(x, y)=Z (x-y)+\epsilon Z (x-\epsilon y)+
\epsilon^2 Z (x-\epsilon^2 y);
\label{V}
\end{equation}
here  $\epsilon=e^{2\pi i/3}$ and $Z''=\zeta$ where $\zeta$ is the Weierstrass zeta-function: $\zeta'=-\wp, \  (\wp')^2=4\wp^3-g_3$.
Degenerations of this solution correspond to
\begin{equation}
H=-\sum_{j\ne i}\frac{\lambda^i-\lambda^j}{3a^2_ia^2_j}\tilde V(a_iu_i, a_ju_j)
\label{32}
\end{equation}
where
$$
\tilde V(x, y)=(x-y)\ln (x-y)+\epsilon(x-\epsilon y)\ln (x-\epsilon y)+
\epsilon^2(x-\epsilon^2 y)\ln (x-\epsilon^2 y),
$$
and
\begin{equation}
H=-\sum_{j\ne i}\frac{\lambda^i-\lambda^j}{a^2_ia^2_j}(a_iu_i-a_ju_j)\ln (a_iu_i-a_ju_j),
\label{33}
\end{equation}
respectively. Further examples include
\begin{equation}
H=\frac{\lambda^1-\lambda^2}{a_2^2} u^2_1 \zeta(a_2u_2)+\frac{\lambda^1-\lambda^3}{a_3^2} u^2_1 \zeta(a_3u_3)-\frac{2}{3}\frac{\lambda^2-\lambda^3}{a_2^2a_3^2}V(a_2u_2, a_3u_3)
\label{331}
\end{equation}
where $V$ is the same as in (\ref{V}). This potential possesses a degeneration
\begin{equation}
H=(\lambda^1-\lambda^2) u^2_1 u^2_2+(\lambda^2-\lambda^3) \zeta(u_3+c) u^2_2-(\lambda^3-\lambda^1) \zeta(u_3) u^2_1,
\label{34}
\end{equation}
here $\zeta'=-\wp, \ (\wp')^2=4\wp^3+4$, and $c$ is the zero of $\wp$ such that $\wp(c)=0, \ \wp'(c)=2$. It possesses a further  quartic degeneration,
\begin{equation}
H= (\lambda^1-\lambda^2) u_1^2 u_2^2+  (\lambda^2-\lambda^3) u_2^2
u_3^2+  (\lambda^3-\lambda^1) u_3^2 u_1^2.
\label{35}
\end{equation}
We have also found the following (non-symmetric) examples:
\begin{equation}
\begin{array}{c}
H=(p u_1+q u_3) \ln{(p u_1+q u_3)}-\frac{1}{6} p
(\lambda^1-\lambda^2) (\lambda^1-\lambda^3) u_1^3-\\
\frac{1}{6} q  (\lambda^3-\lambda^1) (\lambda^3-\lambda^2)
u_3^3+p  (\lambda^3-\lambda^2) u_2 u_3+q  (\lambda^2-\lambda^1)
u_1 u_2,
\label{36}
\end{array}
\end{equation}
\begin{equation}
H=(\lambda^2-\lambda^1) u_2 u_1^2 +(\lambda^2-\lambda^3) u_2 u_3^2+\frac{1}{10}(\lambda^2-\lambda^3)(\lambda^3-\lambda^1)u_3^5+\frac{u_1^2}{u_3},
\label{37}
\end{equation}
and
\begin{equation}
\begin{array}{c}
H=(\lambda^2-\lambda^1) u_2 u_1^2 +(\lambda^2-\lambda^3) u_2 u_3^2+ \\
\frac{p}{15q^2}(\lambda^2-\lambda^1)(\lambda^1-\lambda^3)u_1^5+\frac{q}{15p^2}(\lambda^2-\lambda^3)(\lambda^3-\lambda^1)u_3^5+ u_3 G\Big(\frac{u_1}{u_3} \Big),
\end{array}
\label{38}
\end{equation}
where
$$
G(x)=(p x+ q) \log{(p x+ q)}+\epsilon (p x+\epsilon q) \log{(p
x+\epsilon q)} +\epsilon^2 (p x+\epsilon^2 q) \log{(p x+\epsilon^2
q)}.
$$
Up to the equivalence transformations, the above examples exhaust the  list  of integrable potentials. We claim that all examples appearing in the classification possess dispersionless Lax pairs and an infinity of hydrodynamic reductions (this will be demonstrated in Sect. 5.1--5.2).
\end{theorem}

\medskip

\centerline{\bf Proof of Theorem~\ref{solH123}: }

\medskip

We can assume that all mixed partial derivatives $H_{ij}$ are non-zero.  It follows from
(\ref{H123}) that
\begin{equation}
    \frac{\partial^3}{\partial u_1 \partial u_2 \partial u_3} \left(\frac{H_{12} H_{13}}{H_{23}}\right)=
    \frac{\partial^3}{\partial u_1 \partial u_2 \partial u_3} \left(\frac{H_{12} H_{23}}{H_{13}}\right)=
    \frac{\partial^3}{\partial u_1 \partial u_2 \partial u_3} \left(\frac{H_{13} H_{23}}{H_{12}}\right)=0.
\label{122}
\end{equation}
The further analysis depends on the value of the expression
\begin{equation} \label{123}
   \frac{\partial H_{12}}{\partial u_2}   \frac{\partial H_{23}}{\partial u_3}  \frac{\partial H_{13}}{\partial
   u_1}+
    \frac{\partial H_{12}}{\partial u_1}   \frac{\partial H_{23}}{\partial u_2}  \frac{\partial H_{13}}{\partial u_3},
\end{equation}
which appears as a denominator when solving the equations (\ref{122}).

\noindent {\bf Case I.} The expression (\ref{123}) is nonzero.
In this case  equations (\ref{122}) are equivalent to
$$
 \begin{array}{l}
\displaystyle F_{u_{1},u_{2}}=\frac{K_{u_{1}}}{K} \,F_{u_{2}}+
\frac{G_{u_{2}}}{G} \,F_{u_{1}}-
 \frac{K_{u_{1}}}{K} \frac{G_{u_{2}}}{G} \,F ,
 \\[5mm]
\displaystyle G_{u_{2},u_{3}}=\frac{F_{u_{2}}}{F} \,G_{u_{3}}+
\frac{K_{u_{3}}}{K} \,G_{u_{2}}-
  \frac{F_{u_{2}}}{F} \frac{K_{u_{3}}}{K}  \,G ,
 \\[5mm]
\displaystyle K_{u_{3},u_{1}}=\frac{G_{u_{3}}}{G} \,K_{u_{1}}+
\frac{F_{u_{1}}}{F} \,K_{u_{3}}-
   \frac{G_{u_{3}}}{G} \frac{F_{u_{1}}}{F}  \,K ,
  \end{array}
$$
where $F=1/H_{12},$ $G=1/H_{23},$ $K=1/H_{13}.$  Keeping in mind
that $F_3=G_1=K_2=0$, we can rewrite these equations in the form
\begin{equation}
\left(\frac{F}{GK}\right)_{12}=0, ~~~
\left(\frac{G}{FK}\right)_{23}=0, ~~~
\left(\frac{K}{FG}\right)_{13}=0, ~~~ F_3=G_1=K_2=0. \label{sh}
\end{equation}
The system (\ref{sh}) possesses obvious symmetries
\begin{equation}
\begin{array}{c}
F\to f_1(u_1)f_2(u_2)F, ~~~
G\to f_2(u_2)f_3(u_3)G, ~~~
K\to f_1(u_1)f_3(u_3)K, ~~~ \\
\ \\
u_1\to g_1(u_1), ~~~ u_2\to g_2(u_2), ~~~ u_3\to g_3(u_3);
\end{array}
\label{symm}
\end{equation}
here $f_i$ and $g_i$ are six arbitrary functions of
the indicated arguments. As a first step, we introduce the
new variables
$$
p=\frac{K_1}{K}-\frac{F_1}{F}, ~~~ q=\frac{F_2}{F}-\frac{G_2}{G},
~~~ r=\frac{G_3}{G}-\frac{K_3}{K},
$$
which are nothing but the invariants of the first `half' of the symmetry group (\ref{symm}). In terms of $p, q, r$,
the equations (\ref{sh}) take the form
\begin{equation}
q_1=-p_2=pq, ~~~ r_2=-q_3=qr, ~~~ p_3=-r_1=pr.
\label{pqr}
\end{equation}
This system is straightforward to solve: assuming $p\ne 0$ (the case when $p=q=r=0$ will be a particular case of
the general formula), one has $q=-p_2/p, \ r=p_3/p$, along with the three commuting Monge-Amp\'ere equations for $p$,
\begin{equation}
p_{23}=0, ~~~ ( \ln p)_{12}=p_2, ~~~ (\ln p)_{13}=-p_3.
\label{Monge}
\end{equation}
The integration of the last two equations implies
$p_1/p=p+2\varphi(u_1, u_3)$ and $p_1/p=-p+2\psi(u_1, u_2)$,
respectively. Thus, $p=\psi (u_1, u_2)-\varphi(u_1, u_3)$, and the
substitution back into the above equations gives $\psi_1(u_1,
u_2)-\psi^2(u_1, u_2)=\varphi_1(u_1, u_3)-\varphi^2(u_1, u_3)$.
The separation of variables provides a pair of Riccati equations,
$\psi_1=\psi^2+V(u_1)$ and $\varphi_1=\varphi^2+V(u_1)$. Thus,
$\psi=-[\ln v]_1, \ \varphi=-[\ln \tilde v]_1$, where $v$ and
$\tilde v$ are two arbitrary solutions of the linear ODE
$v_{11}+V(u_1)v=0$. Therefore, we can represent $\psi$ and
$\varphi$ in the form
$$
\psi=-[\ln (q_2(u_2)p_1(u_1)-p_2(u_2)q_1(u_1))]_1, ~~~
\varphi=-[\ln (q_3(u_3)p_1(u_1)-q_1(u_1)p_2(u_2))]_1,
$$
where $p_1(u_1)$ and $q_1(u_1)$ form a basis of solutions of the linear ODE. Introducing
$w_i(u_i)=q_i(u_i)/p_i(u_i)$, one obtains the final formula
$$
p=\psi-\varphi=\frac{w_1'(w_3-w_2)}{(w_2-w_1)(w_3-w_1)},
$$
leading to
$$
q=\frac{w_2'(w_1-w_3)}{(w_2-w_1)(w_2-w_3)}, ~~~ r=\frac{w_3'(w^2-w^1)}{(w_3-w_1)(w_3-w_2)}.
$$
Here $w_i(u_i)$ can be viewed as three arbitrary functions of one argument. The corresponding $F, G, H$ are given by
$$
F=s_1s_2(w_1-w_2), ~~~ G=s_2s_3(w_2-w_3), ~~~ K=s_1s_3(w_3-w_1),
$$
where $s_i(u_i)$ are three extra arbitrary functions. This implies the ansatz
\begin{equation}   \label{bbb1}
H_{12}=\frac{P(u_1)Q(u_2)}{f(u_1)-g(u_2)}, \qquad
 H_{23}=\frac{Q(u_2)R(u_3)}{g(u_2)-h(u_3)},\qquad
 H_{13}=\frac{P(u_1)R(u_3)}{h(u_3)-f(u_1)},
\end{equation}
(with the obvious identification $w_1(u_1)\to f(u_1), \
s_1(u_1)\to 1/P(u_1)$, etc). We have to consider different cases
depending on how many functions among $f, g, h$ are constant.

\noindent {\bf Subcase 1:  $f'=g'=h'=0$}. Without any loss of generality one can assume
$$
  H_{12}={P(u_1)Q(u_2)}, ~  H_{23}={Q(u_2)R(u_3)}, ~ H_{13}={P(u_1)R(u_3)}.
  $$
  Substituting this ansatz into (\ref{H123}) one can show that the functions $P, Q, R$ must necessarily be linear. Up to  the equivalence transformations, this  leads to a unique quartic potential (\ref{35}):
$$
H= (\lambda^1-\lambda^2) u_1^2 u_2^2+  (\lambda^2-\lambda^3) u_2^2
u_3^2+  (\lambda^3-\lambda^1) u_3^2 u_1^2.
$$

\noindent {\bf Subcase 2: $ f'=g'=0$}. Without any loss of generality one can assume  the following ansatz:
\begin{equation}   \label{bbb2}
  H_{12}= P(u_1)Q(u_2) , \qquad
 H_{23}= Q(u_2)h_1(u_3) ,\qquad
 H_{13}= P(u_1)h_2(u_3) .
\end{equation}
The substitution into (\ref{H123}) implies that $P$ and $Q$ must necessarily be linear. Up to  the equivalence transformations, this results in the potential
$$
H=(\lambda^1-\lambda^2) u^2_1 u^2_2+(\lambda^2-\lambda^3) b(u^3) u^2_2+(\lambda^3-\lambda^1) a(u^3) u^2_1,
$$
where the functions $a$ and $b$ satisfy the ODEs
$$
a''=4\frac{a'}{b'}-2, ~~ b''=4\frac{b'}{a'}-2, ~~ a'b'=2(a+b).
$$
The special case $a=b=u^2_3$ brings us back  to the quartic potential from the previous subcase.
The generic solution of these ODEs takes the form
$a(u_3)=-\zeta(u_3), \ b(u_3)=\zeta(u_3+c)$ where $\zeta$ iz the Weierstrass $\zeta$-function, $\zeta'=-\wp, \ (\wp')^2=4\wp^3+4$, and $c$ is the zero of $\wp$ such that $\wp(c)=0, \ \wp'(c)=2$. This is the case (\ref{34}).

\noindent {\bf Subcase 3: $ f'=0$}. The analysis of this case leads to the ansatz
$$
H=(\lambda^1-\lambda^2) u^2_1 a(u_2)+(\lambda^3-\lambda^1) u^2_1 b(u_3)+h(u_2, u_3)
$$
where
$$
a(u_2)=\frac{1}{a_2^2}\zeta(a_2u_2), ~~ b(u_3)=-\frac{1}{a_3^2}\zeta(a_3u_3),
$$
(here $a_2, a_3$ are arbitrary constants), and the second order derivatives of $h(u_2, u_3)$ are given by
$$
H_{23}=4 \frac{\lambda^2-\lambda^3}{a_2a_3}\frac{\wp(a_2u_2)\wp (a_3u_3)}{\wp'(a_2u_2)-\wp'(a_3u_3)},
$$
$$
H_{22}=4\frac{\lambda^2-\lambda^3}{a_3^2}\left(\frac{1}{2}\zeta(a_3u_3) -\frac{\wp^2(a_3u_3)}{\wp'(a_2u_2)-\wp'(a_3u_3)}\right),
$$
$$
H_{33}=4\frac{\lambda^3-\lambda^2}{a_2^2}\left(\frac{1}{2}\zeta(a_2u_2) -\frac{\wp^2(a_2u_2)}{\wp'(a_3u_3)-\wp'(a_2u_2)}\right).
$$
This is the case (\ref{331}).

\noindent {\bf Generic subcase: $f'(x)\, g'(x)\, h'(x) \ne 0.$} From (\ref{bbb1}) and (\ref{H123})  we find all third order derivatives
 of $H.$ The compatibility conditions $\partial_i H_{jji}=\partial_j
 H_{iij}$ give rise to six functional-differential equations for the
 functions $f,g,h,P,Q,R$.
It follows from (\ref{H123}) that
\begin{equation}
\label{extra}
\begin{array}{c}
\partial_{u_{1}} \Big(R_1+(\lambda^{1} - \lambda^{3})
H_{22} + (\lambda^{2} - \lambda^{1}) H_{33} \Big) = 0, \\
\ \\
\partial_{u_{2}}  \Big(R_2+(\lambda^{2} - \lambda^{1})
H_{33} + (\lambda^{3} - \lambda^{2}) H_{11} \Big) = 0, \\
\ \\
\partial_{u_{3}}  \Big(R_3+(\lambda^{3} - \lambda^{2})
H_{11} + (\lambda^{1} - \lambda^{3}) H_{22} \Big) = 0.
\end{array}
\end{equation}
These give us three more equations for $f,g,h,P,Q,R$, so that we
have nine equations altogether. Substituting the values of the
third order derivatives of $H$ into the first equation
(\ref{extra}), taking the numerator and dividing  by the common
factor $P(u_1)^2 Q(u_2)^2 R(u_3),$ we get a fourth degree
polynomial in $f,g,h,P,Q,R$, and  first order derivatives thereof.
Applying to this polynomial the  differential operator
$$
\frac{1}{f'(u_1)\, g'(u_2) }\partial_1 \partial_2 \frac{1}{f'(u_1)\,
g'(u_2)\, h'(u_3)}\partial_1 \partial_2
\partial_3,
$$
we arrive at a separation of variables,
$$
\frac{(\lambda^2-\lambda^3)(P'''(u_1) f'(u_1)-P''(u_1)
f''(u_1))}{f'(u_1)^3}=\frac{(\lambda^3-\lambda^1)(Q'''(u_2)
g'(u_2)-Q''(u_2) g''(u_2))}{g'(u_2)^3}=2 c.
$$
Integrating twice, we obtain
$$
(\lambda^2-\lambda^3) P'=c f^2+a_1 f+b_1, \qquad
(\lambda^3-\lambda^1) Q'=c g^2+a_2 g+b_2.
$$
Analogously,
$$
(\lambda^1-\lambda^2) R'=c h^2+a_3 g+b_3.
$$
Using these relations we eliminate all derivatives of $P,Q$ and $R$ from our
nine equations. As a result, we obtain a  linear  system of nine equations
for the three unknowns $P,Q,R$. This system is consistent (that is, the rank of the extended matrix  is $\leq 3$)  if and only if  $a_1=a_2=a_3=a$,  $b_1=b_2=b_3=b$, and
\begin{equation}
\label{rank}
\begin{array}{c}
4 (c h^2+b h+a) h'^2 f''-4 (c f^2+b f+a) f'^2 h''+\\
(f-h)\Big(2 c (f^2+f h+h^2)+3 b (f+h)+6 a  \Big) f'' h''=0,\\
\ \\
4 (c f^2+b f+a) f'^2 g''-4 (c g^2+b g+a) g'^2 f''+\\
(g-f)\Big(2 c (g^2+g f+f^2)+3 b (g+f)+6 a  \Big) g'' f''=0, \\
\ \\
4 (c g^2+b g+a) g'^2 h''-4 (c h^2+b h+a) h'^2 g''+\\
(h-g)\Big(2 c (h^2+h g+g^2)+3 b (h+g)+6 a  \Big) h'' g''=0.
\end{array}
\end{equation}
Hence, we have either $c=b=a=0$ or $f''=g''=h''=0$, otherwise $f'' g'' h'' \ne 0$.
If  $c=b=a=0$ then the linear system for $P,Q,R$ becomes homogeneous. Its rank equals two if and only if $f''=g''=h''=0.$ In this case
\begin{equation}\label{simp}
 P f' (\lambda^3-\lambda^2)= Q g' (\lambda^1-\lambda^3)=R h' (\lambda^2-\lambda^1)=const.
\end{equation}
If $f''=g''=h''=0$ then the rank of the system  also equals  two. The requirement that the rank of the extended matrix equals two as well leads to $c=b=a=0$. Thus, this case reduces to the previous one.

Suppose now that $f'' g'' h'' \ne 0$.
Solving the linear system for $P, Q, R$ we get
$$
P=\frac{2 (c f^2+b f+a) f'}{(\lambda^2-\lambda^3) f''}, \qquad
Q=\frac{2 (c g^2+b g+a) g'}{(\lambda^3-\lambda^1) g''}, \qquad
R=\frac{2 (c h^2+b h+a) h'}{(\lambda^1-\lambda^2) h''}.
$$
Separating the variables in (\ref{rank}) we ultimately obtain
\begin{equation}
f'^3=c_1 S^2(f), \qquad g'^3=c_2 S^2(g), \qquad h'^3=c_3 S^2(h),
\label{fgh}
\end{equation}
and
$$
P=\frac{(\lambda^1-\lambda^2)(\lambda^1-\lambda^3) S(f)}{2 f'},
\quad Q=\frac{(\lambda^2-\lambda^1)(\lambda^2-\lambda^3) S(g)}{2
g'},\quad R=\frac{(\lambda^3-\lambda^1)(\lambda^3-\lambda^2) S(h)}{2
h'},
$$
where $S(x)$ is a polynomial of degree $\leq 3$, and $c_i$
are arbitrary constants (the polynomial $S(z)$ can be recovered from $(\lambda^1-\lambda^2)(\lambda^1-\lambda^3)(\lambda^2-\lambda^3)S'=6(cz^2+bz+a)$).
Notice that the case (\ref{simp}) is a particular case of the above with $S=const.$

We point out that the right hand sides of (\ref{bbb1}) possess the
following $SL(2, R)$-invariance,
$$
f\to \frac{\alpha f+\beta}{\gamma f+\delta}, ~~ g\to \frac{\alpha g+\beta}{\gamma g+\delta}, ~~
h\to \frac{\alpha h+\beta}{\gamma h+\delta}, ~~~~~
P\to \frac{P}{\gamma f+\delta}, ~~ Q\to \frac{Q}{\gamma g+\delta}, ~~ R\to \frac{R}{\gamma h+\delta},
$$
which can be used to bring the polynomial $S$  to a canonical form. There are three cases to consider.

\noindent {\bf Three distinct roots:}  in this case one can reduce  $S$  to a quadratic, $S(x)=x^2+g_3$, so that the ODEs (\ref{fgh}) imply $f=\wp'(a_1u_1), \ g=\wp'(a_2u_2), \ h=\wp'(a_3u_3)$ where  $27a_i^3=2c_i$ and $\wp$ is the Weierstrass $\wp$-function: $(\wp')^2=4\wp^3-g_3$.  Up to a constant multiple, this leads to
$$
H_{ij}=\frac{\lambda^i-\lambda^j}{a_ia_j}\frac{\wp(a_iu_i)\wp(a_ju_j)}{\wp'(a_iu_i)-\wp'(a_ju_j)}, ~~~
H_{ii}=\sum_{j\ne i}\frac{\lambda^i-\lambda^j}{a_j^2}\left(\frac{1}{2}\zeta(a_ju_j)-\frac{\wp^2(a_ju_j)}{\wp'(a_iu_i)-\wp'(a_ju_j)}\right).
$$
The corresponding potential $H({\bf u})$ is given by (\ref{31}).

\noindent {\bf Double root:}  in this case one can assume $S(x)=x$, so that the
ODEs (\ref{fgh}) imply $f=(a_1u_1)^3, \ g=(a_2u_2)^3, \ h=(a_3u_3)^3$, here $27a_i^3=c_i$.
Up to a constant multiple, this leads to
$$
H_{ij}=\frac{(\lambda^i-\lambda^j)u_iu_j}{(a_iu_i)^3-(a_ju_j)^3}, ~~~
H_{ii}=-\sum_{j\ne i}\frac{(\lambda^i-\lambda^j)u_j^2}{(a_iu_i)^3-(a_ju_j)^3}.
$$
The corresponding potential $H({\bf u})$ is given by (\ref{32}).

\noindent {\bf Triple root:} in this case  $S$  can be reduced to $S=1$, so that the ODEs
(\ref{fgh}) imply $f=a_1u_1, \ g=a_2u_2, \ h=a_3u_3$, here $a_i^3=c_i$. Up to a constant multiple, this leads to
\begin{equation}
\label{tripsol}
H_{ij}=\frac{\lambda^i-\lambda^j}{a_ia_j(a_iu_i-a_ju_j)}, ~~~
H_{ii}=-\sum_{j\ne
i}\frac{\lambda^i-\lambda^j}{a^2_j(a_iu_i-a_ju_j)},
\end{equation}
and the corresponding potential $H({\bf u})$ is given by (\ref{33}).
Notice, however, that for this potential the expression (\ref{123}) equals zero. Formally, it should be considered as an example from the Case II below.

\noindent {\bf Case II}. This is the case when the expression (\ref{123}) equals zero, although both terms in (\ref{123}) are nonzero:
 \begin{equation}   \label{det1}
    H_{122}H_{233} H_{113}=-
   H_{112}H_{223} H_{133}
     \ne 0;
 \end{equation}
 an integrable example from this class is provided by
\begin{equation}
\label{admissible}
H_{ij}=\frac{\lambda^i-\lambda^j}{a_ia_j(a_iu_i-a_ju_j)};
\end{equation}
it appears in the `triple root' case above. A detailed analysis below shows that this case possesses no other non-trivial solutions. Rewriting (\ref{det1}) in the form
$$
\frac{ H_{122}}{H_{112}}\frac{H_{233}}{H_{223}}\frac{ H_{113}}{
 H_{133}}=-1
 $$
one can set
$$
\frac{ H_{122}}{H_{112}}=-\frac{l(u^1)}{m(u^2)}, ~~
\frac{ H_{233}}{H_{223}}=-\frac{m(u^2)}{n(u^3)}, ~~
\frac{ H_{113}}{H_{133}}=-\frac{n(u^3)}{l(u^1)}.
$$
Thus,
\begin{equation}
\label{Hij}
 H_{12}=\frac{1}{P(x)}, \qquad  H_{23}=\frac{1}{Q(y)},
 \qquad     H_{13}=\frac{1}{R(z)},
\end{equation}
 where $x=\alpha(u_1)-\beta(u_2),$ $y=\beta(u_2)-\gamma(u_3)$ and
 $z=-x-y$ for
 some functions $\alpha, \beta, \gamma$ such
 that $\alpha'=1/l, \ \beta'=1/m, \  \gamma'=1/n.$
Substituting (\ref{Hij}) into (\ref{H123}) and integrating once, one gets
\begin{gather}
\label{HiiPQR}
\begin{aligned}
H_{11} = \frac{(\lambda^{1}-\lambda^{2}) P^{2} +
(\lambda^{3}-\lambda^{1}) R^{2}}{(\lambda^{2}-\lambda^{3}) P Q R}+
\mu(u_{2},u_{3}),
\\
H_{22} = \frac{(\lambda^{1}-\lambda^{2}) P^{2} +
(\lambda^{2}-\lambda^{3}) Q^{2}}{(\lambda^{3}-\lambda^{1}) P Q
R}+\nu(u_{1},u_{3}),
\\
H_{33} = \frac{(\lambda^{3}-\lambda^{2}) Q^{2} +
(\lambda^{3}-\lambda^{1}) R^{2}}{(\lambda^{1}-\lambda^{2}) P Q R}+
\eta(u_{1},u_{2}).
\end{aligned}
\end{gather}
Expressing  six partial derivatives of the functions
$\mu(u_{2},u_{3})$, $\nu(u_{1},u_{3})$, $\eta(u_{1},u_{2})$ from
the six compatibility conditions $\partial_{j}H_{ii} =
\partial_{i}H_{ij}$, and
substituting them into the equations $\partial_{1}J
=0$ and $\partial_{2}J =0$, we obtain
\begin{gather}
\label{w1w2}
\begin{aligned}
&w_{1} \alpha'(u_{1}) + w_{2} \gamma'(u_{3})=0, \qquad &w_{3}
\beta'(u_{2}) + w_{4} \gamma'(u_{3})=0,\\
&\partial_{2}(w_{1}) \alpha'(u_{1}) + \partial_{2}(w_{2})
 \gamma'(u_{3})=0, \qquad
&\partial_{1}(w_{3}) \beta'(u_{2}) + \partial_{1}(w_{4})
 \gamma'(u_{3})=0,
\end{aligned}
\end{gather}
where
\begin{align*}
w_{1} &= (\lambda^{2} - \lambda^{3}) Q (R P' Q' + Q P' R' - P Q'
R'),\qquad w_{2} = (\lambda^{2} - \lambda^{1}) P (R P' Q' - Q P'
R' + P Q'
R'),\\
w_{3} &= (\lambda^{1} - \lambda^{3}) R (R P' Q' + Q P' R' - P Q'
R'),\qquad w_{4} = (\lambda^{2} - \lambda^{1}) P (R P' Q' - Q P'
R' - P Q' R').
\end{align*}
The equations~(\ref{w1w2})$_{2}$  are obtained from (\ref{w1w2})$_{1}$ upon differentiation  by
$u_{2}$ and  $u_{1}$, respectively. Since $\alpha' , \beta' $ and $\gamma'$ are nonzero, the system~(\ref{w1w2}) is consistent iff $P$, $Q$ and $R$
satisfy the following conditions:
\begin{equation}
\label{detPQR} w_{1} \partial_{2}(w_{2}) - w_{2}
\partial_{2}(w_{1}) = 0,\qquad w_{3}
\partial_{1}(w_{4}) - w_{4} \partial_{1}(w_{3}) = 0.
\end{equation}
Let us observe that the equations~(\ref{122}) (which also hold
in this case)  can be rewritten as follows:
\begin{equation}
\label{CaseIISist1} \frac{p'}{p} = \frac{p q}{r} - \frac{r q}{p} +
\frac{r'}{r}, \qquad \frac{q'}{q} = \frac{p q}{r} - \frac{r p}{q}
+ \frac{r'}{r},\qquad \frac{p'}{p} = \frac{r p}{q} - \frac{r q}{p}
+ \frac{q'}{q};
\end{equation}
here we use the notation $p =
P'/P$, $q = Q'/Q$, $r = R'/R$, and prime
denotes  the derivative of  functions with respect to their
arguments. Note that only two of the above equations are independent.
Differentiating, for instance, the first two equations
in~(\ref{CaseIISist1}) by $x$ and $y$
 and eliminating $r$ and $r'$, one ends up at the following
 relations involving $p$ and $q$:
\begin{gather}
\label{CaseIISist2}
\begin{aligned}
&\left[\left(\frac{p'}{p}\right)'+ \left(\frac{q'}{q}\right)'
\right] \left(p^{2} - q^{2} \right) =
\left[\left(\frac{p'}{p}\right)^{2} - \left(\frac{q'}{q}
\right)^{2} \right] \left(p^{2} + q^{2} \right), \\
&\left[\left(\frac{p'}{p}\right)^{2} - \left(\frac{q'}{q}
\right)^{2} \right] = 2 \left(p^{2}-q^{2} \right)-
\left[\left(\frac{p'}{p}\right)'- \left(\frac{q'}{q}\right)'
\right].
\end{aligned}
\end{gather}
Similarly, one can eliminate $p$  and $p'$ obtaining
the analog of~(\ref{CaseIISist2}) for $q$ and $r$. All these
relations imply
\begin{equation}
\label{sepPQ} p^{2} \left(\frac{p'}{p} \right)' - p^{4} =
q^{2}\left(\frac{q'}{q} \right)'- q^{4} = r^{2}\left(\frac{r'}{r}
\right)'- r^{4}.
\end{equation}
Thus, $p$, $q$ and $r$ must satisfy an ODE of the form
$$
f f'' - (f')^{2} - f^{4} + k = 0
$$
where $f = f(\zeta)$ and $k$ is an arbitrary constant. If $k = 0$
then
$$
f = \frac{\nu}{\sin \nu \zeta} ~~~ {\rm or} ~~~ f = \frac{1}{\zeta}
$$
where $\nu$ is an arbitrary constant. Note that the second
solution is a limit of the first as $\nu \to 0$. If $k \neq 0$
we have
$$
f = \frac{k^{\frac{1}{4}}}{\tanh k^{\frac{1}{4}} \zeta}   ~~~ {\rm or} ~~~     f =
\frac{1}{\nu} sn\left(\nu \sqrt{k} \zeta; \frac{1}{\nu^{4} k}
\right)
$$
where $sn$ is the Jacobi elliptic sine function: $(sn')^{2} =
(1-sn^{2}) \left(1 - 1/(\nu^{4} k) sn^{2}\right)$. Using $p =
P'/P$ and integrating once, we obtain
\begin{align}
\label{PQR1} P &= c_{3} x,  \qquad P = c_{3} \frac{\tan{\nu x
}}{\nu},\qquad P = c_{3}  \frac{\sinh{k^{\frac{1}{4}}
x}}{k^{\frac{1}{4}}}, \qquad {\rm or} \qquad
 P = c_{3} dn- c_{3} \frac{cn}{\nu^{2} \sqrt{k}}\;.
\end{align}
Analogously, $Q$ and $R$  can be obtained from these formulae by
cycling the indices $c_3 \to c_1 \to c_2$ and the variables $x \to
y \to z$. Here $cn$ and $dn$ are the Jacobi elliptic functions
$cn(\nu \sqrt{k} x; 1/(\nu^{4} k))$ and $dn(\nu \sqrt{k} x;
1/(\nu^{4} k))$, and $c_{1}$, $c_{2}$ and $c_{3}$ are arbitrary
constants. It turns out that only  linear and trigonometric
solutions in~(\ref{PQR1}) satisfy the condition~(\ref{detPQR}).
Thus,  hyperbolic and  elliptic solutions can be dropped. The
substitution of the linear solution into one of the
equations~(\ref{w1w2})$_{1}$ implies that  the functions $\alpha$,
$\beta$ and $\gamma$ must be linear. One recovers the
solution~(\ref{tripsol}) by setting $c_{1} = a_{2}
a_{3}/(\lambda^{2}-\lambda^{3})$, $c_{2} = a_{1}
a_{3}/(\lambda^{3}-\lambda^{1})$,  $c_{3} = a_{1}
a_{2}/(\lambda^{1}-\lambda^{2})$. Finally, the substitution of the
trigonometric solution~(\ref{PQR1}) also implies that $\alpha$,
$\beta$ and $\gamma$ must be linear, however, the compatibility
conditions for the systems~(\ref{Hij}) and~(\ref{HiiPQR}) are not satisfied.

\noindent {\bf Case III.} This is the case when both terms in (\ref{123}) equal zero separately:
  \begin{equation}   \label{det2}
     \frac{\partial H_{12}}{\partial u_2}   \frac{\partial H_{23}}{\partial u_3}  \frac{\partial H_{13}}{\partial
     u_1}=
      \frac{\partial H_{12}}{\partial u_1}   \frac{\partial H_{23}}{\partial u_2}  \frac{\partial H_{13}}{\partial u_3}
      = 0.
  \end{equation}
Up to permutations of indices, we have to consider the following three
subcases.

\noindent {\bf Subcase 1:} $H_{12}=const\ne 0$. It follows from (\ref{H123}) that
$$
(\lambda^1-\lambda^2) H_{13} H_{233}=(\lambda^3-\lambda^1) H_{12}
H_{223}, \qquad (\lambda^2-\lambda^1) H_{23}
H_{133}=(\lambda^3-\lambda^2) H_{12} H_{113}.
$$
Differentiating the first equation with respect to $u_1$ we obtain
$H_{113} H_{233}=0$. If $H_{233}=0$ then the first equation implies
$H_{23}=const$. Otherwise, it follows from the second equation that
$H_{13}=const.$ Without any loss of generality we assume that
$H_{23}=const\ne 0$. Setting $H_{12}=q(\lambda^2-\lambda^1), \ H_{23}=p(\lambda^3-\lambda^2)$, $p, q=const$,  and substituting  into (\ref{H123}) one arrives, up to the equivalence transformations, at the following potential $H$:
$$
H(u_1,u_2,u_3)=(p u_1+q u_3) \ln{(p u_1+q u_3)}-\frac{1}{6} p
(\lambda^1-\lambda^2) (\lambda^1-\lambda^3) u_1^3-
$$
$$
-\frac{1}{6} q  (\lambda^3-\lambda^1) (\lambda^3-\lambda^2)
u_3^3+p  (\lambda^3-\lambda^2) u_2 u_3+q  (\lambda^2-\lambda^1)
u_1 u_2.
$$

\noindent{\bf Subcase 2:} $H_{12}=f(u_1)$, $H_{23}=g(u_3)$. One can prove that in this case
$$
H(u_1,u_2,u_3)=\alpha u_1^2 u_2+\beta u_2 u_3^2+\gamma u_1^5+\delta
u_3^5+ u_3 G\Big(\frac{u_1}{u_3} \Big)
$$
for some constants $\alpha,\beta,\gamma,\delta$.
The function $G$ has to satisfy an equation of the form
$$
G''(x)=\frac{z_1}{z_2+z_3 x^3},
$$
where $z_i$ are some constants. If $z_3=0$ we have $G(x)=x^2$. In
this case
$$
\alpha= (\lambda^2-\lambda^1), \qquad \beta=
(\lambda^2-\lambda^3), \qquad \gamma=0,\qquad \delta=\frac{1}{10}
(\lambda^2-\lambda^3)(\lambda^3-\lambda^1),
$$
which gives (\ref{37}). The case $z_2=0$ is equivalent to the above. Otherwise,
$$
G(x)=(p x+ q) \log{(p x+ q)}+\epsilon (p x+\epsilon q) \log{(p
x+\epsilon q)} +\epsilon^2 (p x+\epsilon^2 q) \log{(p x+\epsilon^2
q)}.
$$
In this case
$$
\alpha= (\lambda^2-\lambda^1), \quad \beta= (\lambda^2-\lambda^3),
\quad \gamma=\frac{p }{15 q^2}
(\lambda^2-\lambda^1)(\lambda^1-\lambda^3),\quad \delta=\frac{q
}{15 p^2} (\lambda^2-\lambda^3)(\lambda^3-\lambda^1),
$$
which gives (\ref{38}).

\noindent{\bf Subcase 3:} $H_{12}=f(u_1)$, $H_{13}=g(u_1)$.  A direct calculation shows that this case gives no non-trivial examples.

\noindent This finishes the proof of Theorem 3.

\subsection{Dispersionless Lax pairs}

In this section we prove  that the diagonalizability conditions (\ref{H123}) imply  the existence of  the dispersionless Lax pairs
(Theorem 4), and explicitly calculate  Lax pairs for some of the most `symmetric' examples appearing in the classification list of Theorem 3.

\medskip

\noindent {\bf Example 1.} Let us consider the quartic potential (\ref{35}),
$$
H= (\lambda^1-\lambda^2) u_1^2 u_2^2+  (\lambda^2-\lambda^3) u_2^2
u_3^2+  (\lambda^3-\lambda^1) u_3^2 u_1^2,
$$
which is a three-component generalization of the potential (\ref{gen2}) from Theorem 1 (we have verified that this example possesses no natural four-component extensions). The corresponding system (\ref{Hessian}) has a Lax pair
$$
\psi_T=\lambda_1 a_1(\xi) u_1^2+\lambda_2 a_2(\xi) u_2^2+\lambda_3
a_3(\xi) u_3^2, \qquad \psi_X=-a_1(\xi) u_1^2-a_2(\xi) u_2^2-a_3(\xi)
u_3^2;
$$
here $\xi=\psi_Y$ and the functions $a_i(\xi)$ satisfy the ODEs
$$
a_1'=\frac{4 a_1}{a_3}+2, \qquad a_2'=\frac{4 a_2}{a_1}+2, \qquad
a_3'=\frac{4 a_3}{a_2}+2, \qquad a_1 a_2+a_2 a_3+a_3 a_1=0.
$$
Equivalently,
$$
a_3=\frac{4 a_1}{a_1'-2}, \qquad a_2=-\frac{4 a_1}{a_1'+2}, \qquad 2
a_1 a_1''=3 a_1'^2-12.
$$
Without any loss of generality one can set
$$
a_1=\wp(\xi), ~~ a_2= \wp(\xi+c), ~~ a_3=\wp(\xi-c)
$$
where $\wp$ is the Weierstrass $\wp$-function, $(\wp')^2=4\wp^3+4$, and $c$ is the zero of $\wp$ such that $\wp(c)=0, \ \wp'(c)=2$.

\medskip

\noindent {\bf Example 2.} Let us consider the potential (\ref{33}),
$$
H=-\sum_{j\ne i}\frac{\lambda^i-\lambda^j}{a^2_ia^2_j}(a_iu_i-a_ju_j)\ln (a_iu_i-a_ju_j),
$$
which is a three-component generalization of the potential (\ref{gen11}) from Theorem 1.
The corresponding system (\ref{Hessian}) possesses the Lax pair
$$
\psi_T=-\sum \frac{\lambda^i}{a_i^2} \ln (a_i u_i-\psi_Y), ~~~
 \psi_X= \sum \frac{1}{a_i^2} \ln (a_i u_i-\psi_Y).
$$
This Lax pair appeared previously in \cite{Odesskii}.

\medskip

\noindent {\bf Example 3.} Let us consider the potential (\ref{31}),
$$
H=-\sum_{j\ne i}\frac{\lambda^i-\lambda^j}{6a^2_ia^2_j}V(a_iu_i, a_ju_j).
$$
One can show that the  corresponding system (\ref{Hessian}) has the Lax pair
$$
\psi_T=-\sum \frac{\lambda^i}{a_i^2} f(a_i u_i, \psi_Y), ~~~
 \psi_X= \sum \frac{1}{a_i^2} f (a_i u_i, \psi_Y)
$$
where the dependence of  $f(u, \xi)$ on its arguments  (here $\xi=\psi_Y$) is governed by
$$
f_u=\frac{\wp(u)\wp(\xi)}{\wp'(u)-\wp'(\xi)}, ~~~ f_{\xi}= \frac{\wp^2(u)}{\wp'(\xi)-\wp'(u)}-\frac{1}{2}\zeta(u).
$$
Explicitly, one has
$$
f(u, \xi)=\frac{1}{6}\ln \sigma(u-\xi)+\frac{\epsilon}{6} \ln \sigma(\epsilon u-\xi)+\frac{\epsilon^2}{6}
\ln \sigma(\epsilon^2u-\xi),
$$
here $\sigma$ is the Weierstrass sigma-function: $\sigma'/\sigma=\zeta$. In a different parametrization,  this Lax pair appeared   in \cite{Odesskii} in the classification of dispersionless Lax pairs with movable singularities. We point out that both examples 2 and 3 generalize to $n$-component case in a straightforward way (allowing the summation to go from $1$ to $n$).

\medskip

\noindent In fact, the following general  result holds:

\begin{theorem}
\label{theoLax} Any system~\textup{(\ref{Hessian})}
satisfying the  diagonalizability conditions (\ref{H123})
possesses a dispersionless Lax pair.
\end{theorem}

\medskip

\centerline{\bf Proof: }

\medskip

\noindent We look for a Lax pair in
the form~(\ref{Lax2}). The compatibility condition $\psi_{tx} =
\psi_{xt}$ results in the following set of relations:
\begin{gather}
\label{pseudo1}
\begin{aligned}
f_{1} &= \lambda^{1} g_{1}, ~~ f_{2} &= \lambda^{2} g_{2}, ~~
f_{3} &= \lambda^{3} g_{3}, \\
\end{aligned}
\end{gather}
and
\begin{gather}
\begin{aligned}
\label{pseudo2}
f_{p} g_{1}& = H_{11} g_{1} + H_{21} g_{2} + H_{31} g_{3} + g_{p} f_{1}, \\
f_{p} g_{2} &= H_{12} g_{1} + H_{22} g_{2} + H_{32} g_{3} + g_{p} f_{2}, \\
f_{p} g_{3} &= H_{13} g_{1} + H_{23} g_{2} + H_{33} g_{3} + g_{p} f_{3}, \\
\end{aligned}
\end{gather}
where we have set $p = \psi_{y}$, $f_i=\partial_i f,$ and
$g_i=\partial_i g$. The relations~(\ref{pseudo1})
and~(\ref{pseudo2}) are equivalent to (\ref{Lax1}). Eliminating
$f_{p}$ and $g_{p}$ from (\ref{pseudo2}), one obtains a
single algebraic constraint among the components  $g_1, g_2, g_3$,
which coincides with the left characteristic cone~(\ref{cone}).
The expressions for $f_{p}$ and $g_{p}$ obtained from the first
two equations (\ref{pseudo2}) take the form
\begin{gather}
\label{pseudo3}
\begin{aligned}
f_{p} &= \frac{\left(H_{11} g_{1}  + H_{12} g_{2} + H_{13} g_{3}\right) \lambda^2g_{2} - \left(H_{12} g_{1} + H_{22} g_{2} + H_{23} g_{3} \right) \lambda^1g_{1}}{g_{1} g_{2}(\lambda^2-\lambda^1)}, \\
g_{p} &= \frac{\left(H_{11} g_{1}  + H_{12} g_{2} + H_{13}
g_{3}\right) g_{2} - \left(H_{12} g_{1} + H_{22} g_{2} + H_{23}
g_{3} \right) g_{1}}{g_{1} g_{2}(\lambda^2-\lambda^1)}.
\end{aligned}
\end{gather}
Using the compatibility conditions $f_{ij}=f_{ji}$ and
$f_{ip}=f_{pi}$, we can  express all second order derivatives of
$g$ in the form
\begin{gather}
\label{gij}
\begin{aligned}
g_{12} =  & g_{13}=g_{23}=0,  \\
\ \\
g_{11} = & \frac{g_{1} \left(H_{111} g_{1} + H_{112} g_{2} +
H_{113} g_{3}
 \right)}{H_{12} g_{2} + H_{13} g_{3}}, \\
 \ \\
g_{22} = & \frac{g_{2} \left(H_{221} g_{1} + H_{222} g_{2} +
H_{223} g_{3}
 \right)}{H_{12} g_{1} + H_{23} g_{3}}, \\
 \ \\
g_{33} = & \frac{g_{1} \left(H_{123} g_{1} + H_{223} g_{2} +
H_{332} g_{3}
 \right) \left(\lambda^{3} - \lambda^{1} \right)}
 {H_{13} g_{2} \left(\lambda^{3} - \lambda^{2} \right) + H_{23} g_{1}
 \left(\lambda^{1} - \lambda^{3} \right)}\\
 & +
 \frac{
 g_{2} \left(H_{113} g_{1} + H_{123} g_{2} + H_{331} g_{3} \right) \left(
 \lambda^{2} - \lambda^{3} \right)
 }{H_{13} g_{2} \left(\lambda^{3} - \lambda^{2} \right) + H_{23} g_{1}
 \left(\lambda^{1} - \lambda^{3} \right)}.
\end{aligned}
\end{gather}
It was already mentioned that the condition $J=0$ implies the
decomposition of the left characteristic cone~(\ref{cone}) into
 a linear and  quadratic factors, see (\ref{decomp}). We will assume that $g_1, g_2, g_3$ lie on the  quadratic branch,
\begin{equation}
\label{quadbranch}\Gamma=H_{13} H_{23} (\lambda^{1} - \lambda^{2}) g_{1} g_{2}
+H_{12} H_{23} (\lambda^{3} - \lambda^{1}) g_{1} g_{3} + H_{12}
H_{13} (\lambda^{2} - \lambda^{3}) g_{2} g_{3} =0.
\end{equation}
One can verify that the differential consequences
\begin{equation}
\label{diffGamma} \frac{\partial \Gamma}{\partial {u_{1}}} = 0,
\qquad \frac{\partial \Gamma}{\partial {u_{2}}} = 0,\qquad
\frac{\partial \Gamma}{\partial {u_{3}}} = 0, \qquad
\frac{\partial \Gamma}{\partial {p}} = 0
\end{equation}
hold identically  modulo (\ref{gij}), (\ref{quadbranch})
and (\ref{H123}). Finally, using
computer algebra, it is straightforward to verify that the
consistency conditions for the system~(\ref{gij}) are satisfied identically
modulo~(\ref{quadbranch}) and (\ref{H123}). This completes the
proof of  theorem~\ref{theoLax}.

\subsection{Hydrodynamic reductions}

The aim of this section is to prove that all examples listed in Theorem 3 possess infinitely many $n$-component hydrodynamic reductions parametrized by $n$ arbitrary functions of a single variable. To do so one has to demonstrate the consistency of the relations (\ref{reduction}), (\ref{comm}) where the characteristic speeds $\nu^i$ and $\mu^i$ satisfy the dispersion relation $det(\nu I_3+\mu A+B)=0$, and $\partial_i{\bf u}$ is the right eigenvector of the matrix $\nu^i I_3+\mu^i A+B$ --- see Sect. 2.

\begin{theorem} The diagonalizability conditions (\ref{H123}) are necessary and sufficient for the existence of an infinity of $n$-component hydrodynamic reductions parametrized by $n$ arbitrary functions of a single variable.
\end{theorem}

\medskip

\centerline {\bf Proof:}

\medskip

The necessity follows from the general result of \cite{Fer2} which states that, for a quasilinear system (\ref{2+1}),  the diagonalizability is a necessary condition for the existence of an infinity of hydrodynamic reductions.

The first step to demonstrate the sufficiency is to explicitly parametrize the dispersion curve (\ref{disp}),
which we know to be a {\it rational} curve of degree three (see the Remark after Theorem 2). This can be done as follows.
Let us first calculate the singular point $\nu_0, \mu_0$ on the dispersion curve.  It corresponds to
the situation when the rank of the matrix $\nu I_3+\mu A+B$ drops to one. The associated left eigenvectors constitute a two-dimensional plane given by the first factor in the equation of the left characteristic cone (\ref{decomp}). A simple calculation shows that $\nu_0$ and $\mu_0$ can be obtained from the linear system
$$
\nu_0+\lambda^1\mu_0=\frac{H_{12}H_{13}}{H_{23}}-H_{11},
$$
$$
\nu_0+\lambda^2\mu_0=\frac{H_{12}H_{23}}{H_{13}}-H_{22},
$$
$$
\nu_0+\lambda^3\mu_0=\frac{H_{13}H_{23}}{H_{12}}-H_{33};
$$
notice that these three relations are linearly dependent, indeed, multiplying the first by $\lambda^2-\lambda^3$,  the second by $\lambda^3-\lambda^1$, the third by $\lambda^1-\lambda^2$ and adding them together, one obtains  $J=0$, see (\ref{H123}).
Next, we parametrize the quadratic branch of the left characteristic cone (\ref{decomp}) in the form
$$
g_1=\frac{1}{(\lambda^1+s)H_{23}}, ~~ g_2=\frac{1}{(\lambda^2+s)H_{13}}, ~~ g_3=\frac{1}{(\lambda^3+s)H_{12}},
$$
here $s$ is a parameter. The corresponding relation (\ref{dual}) is equivalent to
$$
\nu+\mu \lambda^1+H_{11}+\frac{\lambda^1+s}{\lambda^2+s}\frac{H_{12}H_{23}}{H_{13}}+
\frac{\lambda^1+s}{\lambda^3+s}\frac{H_{13}H_{23}}{H_{12}}=0,
$$
$$
\nu+\mu \lambda^2+H_{22}+\frac{\lambda^2+s}{\lambda^1+s}\frac{H_{12}H_{13}}{H_{23}}+
\frac{\lambda^2+s}{\lambda^3+s}\frac{H_{13}H_{23}}{H_{12}}=0,
$$
$$
\nu+\mu \lambda^3+H_{33}+\frac{\lambda^3+s}{\lambda^1+s}\frac{H_{12}H_{13}}{H_{23}}+
\frac{\lambda^3+s}{\lambda^2+s}\frac{H_{12}H_{23}}{H_{13}}=0;
$$
we point out that these three relations are also linearly dependent. Solving them for $\nu(s)$ and $\mu(s)$ one obtains a rational parametrization of the dispersion curve:
$$
\nu(s)=\nu_0-\frac{s}{\lambda^1+s}\frac{H_{12}H_{13}}{H_{23}}
-\frac{s}{\lambda^2+s}\frac{H_{12}H_{23}}{H_{13}}
-\frac{s}{\lambda^3+s}\frac{H_{13}H_{23}}{H_{12}},
$$
$$
\mu(s)=\mu_0-\frac{1}{\lambda^1+s}\frac{H_{12}H_{13}}{H_{23}}
-\frac{1}{\lambda^2+s}\frac{H_{12}H_{23}}{H_{13}}
-\frac{1}{\lambda^3+s}\frac{H_{13}H_{23}}{H_{12}};
$$
here $\nu_0$ and $\mu_0$ are  coordinates of the singular point.
 Thus,  the characteristic speeds $\nu^i({\bf R})$ and $\mu^i({\bf R})$ can be represented in the form
\begin{equation}
\nu^i({\bf R})=\nu(s^i), ~~~ \mu^i({\bf R})=\mu(s^i)
\label{rat}
\end{equation}
where $s^i$, which are the parameter values of $n$ points on the dispersion curve, are certain functions of the Riemann invariants:  $s^i=s^i({\bf R})$.
Since in our case the matrix $\nu I_3+\mu A+B$ is symmetric, the left characteristic cone coincides with the right characteristic cone. Thus, the right eigenvector corresponding to the point $\nu^i, \mu^i$ on the dispersion curve is
$$
\left(\frac{1}{(\lambda^1+s^i)H_{23}}, \ \frac{1}{(\lambda^2+s^i)H_{13}}, \ \frac{1}{(\lambda^3+s^i)H_{12}}
\right)^t,
$$
and the relations (\ref{reduction}) take the form
\begin{equation}
\partial_iu_2=\frac{\lambda^1+s^i}{\lambda^2+s^i}\frac{H_{23}}{H_{13} }\ \partial_iu_1, ~~
\partial_iu_3=\frac{\lambda^1+s^i}{\lambda^3+s^i}\frac{H_{23}}{H_{12}} \ \partial_iu_1.
\label{u23}
\end{equation}
Substituting (\ref{rat}) into the commutativity conditions (\ref{comm}) and using (\ref{u23}) one obtains
the relations
\begin{equation}
\partial_j s^i=(...)\partial_ju_1
\label{sij}
\end{equation}
$i\ne j$, where dots denote certain {\it rational} expression in $s^i, s^j$
whose coefficients depend on the second and third order derivatives of the potential $H$.
For example, in the case of the quartic potential (\ref{35}) these relations take the form
$$
\partial_j s^i=\frac{3 (\lambda^1+s^i)(\lambda^2+s^i)(\lambda^3+s^i)(\lambda^1+s^j)}
{(\lambda^1-\lambda^2)(\lambda^1-\lambda^3) (s^j-s^i) \, u_1}
\, \partial_ju_1.
$$
By virtue of (\ref{u23}) and (\ref{H123}), the consistency
conditions  $\partial_j\partial_i u_2=\partial_i\partial_j u_2$ and
$\partial_j\partial_i u_3=\partial_i\partial_j u_3$ imply one and the same relation
\begin{equation}
\partial_i\partial_j u_1=(...)\partial_iu_1\partial_j u_1,
\label{u1ij}
\end{equation}
$i\ne j$, where, again, dots denote a  {\it rational} expression in $s^i, s^j$ whose coefficients depend
on the second and third order derivatives of  $H$.  In the case (\ref{35}), we have
$$\partial_i\partial_j u_1=\frac{Y(s^i,s^j)}{u_1}\,\partial_iu_1\partial_j u_1, $$
where
$$
 Y(\alpha,\beta)=\frac{6 \alpha^2 \beta^2+k_1 (\alpha^2 \beta+\alpha \beta^2)+
 k_2 (\alpha^2+4 \alpha \beta+\beta^2)+k_3 (\alpha+\beta)+k_4}
 {(\lambda^1-\lambda^2)(\lambda^1-\lambda^3)(\alpha-\beta)^2},
$$

$$
k_1=  3 (\lambda^2+\lambda^3+2 \lambda^1), \qquad k_2= (\lambda^1)^2+2 \lambda^1 \lambda^2+ 2 \lambda^1 \lambda^3+
\lambda^2 \lambda^3, $$$$ k_3=3 \lambda^1 (\lambda^1 \lambda^2+ \lambda^1 \lambda^3+2 \lambda^2 \lambda^3),\qquad
k_4=6 (\lambda^1)^2 \lambda^2 \lambda^3.
$$
The relations (\ref{sij}) and (\ref{u1ij}) constitute the so-called Gibbons-Tsarev-type equations
which govern hydrodynamic reductions of the system (\ref{Hessian}).
The last step is to verify  their consistency,
namely, $\partial_k \partial_j s^i=\partial_j \partial_k s^i$ and
$\partial_i\partial_j \partial_k u_1=\partial_i\partial_k\partial_j u_1$
(without any loss of generality one can set $i=1,\  j=2,\  s=3$). If these consistency
conditions are satisfied identically, the  system (\ref{sij}), (\ref{u1ij}) will be in involution,
with the general solution depending on $2n$ arbitrary functions of a single variable.
Up to reparametrizations  $R^i\to f^i(R^i)$ this gives an infinity of hydrodynamic reductions
depending on $n$ arbitrary functions.

We have verified the consistency for all examples appearing in Theorem 3. In fact, rather than considering them case-by-case, one can give a unified proof of the consistency using only the diagonalizability conditions (\ref{H123}). To do so one needs to bring the  system (\ref{H123}) into a passive form. It turns out that all higher order partial derivatives of the potential $H$  can be expressed
in terms of the second order derivatives $H_{ij}$ and the $4$ third order derivatives, say, $H_{122}, H_{113}, H_{223}, H_{233}.$
Second order derivatives are constrained by a single algebraic equation $J=0$, while the values of $H$ and its first order derivatives $H_i$
are arbitrary. This calculation shows that the generic solution of the system (\ref{H123})  should depend on 13
arbitrary constants, which is in full accordance with the results of Section 5. The computation of the expressions
(\ref{sij}) and (\ref{u1ij}), as well as the verification of the consistency conditions have been performed modulo this
passive form. This means that all partial derivatives of $H$ except the basic ones were eliminated, and the basic
derivatives were considered as independent variables related by a single algebraic equation $J=0$. An intense computer
calculation shows that all compatibility conditions are identities in the basic derivatives.

\section{Hamiltonian systems in $3+1$ dimensions}

In this section we establish a number of non-existence results for integrable Hamiltonian systems of hydrodynamic type in $3+1$ dimensions. We will begin with a two-component case. According to the results of \cite{Mokhov1}, there exists a unique two-component Hamiltonian operator of hydrodynamic type which is essentially three-dimensional. Up to a linear transformation of the independent variables it can be cast into a canonical form
$$
P= \left(
\begin{array}{cc}
d/dx&0\\
0 & d/dy
\end{array}
\right)+ \left(
\begin{array}{cc}
0&d/dz\\
d/dz & 0
\end{array}
\right).
$$
The corresponding Hamiltonian systems ${\bf u}_t+P(h_{\bf u})=0$ take the form
$$
u^1_t+(h_1)_x+(h_2)_z=0, ~~~ u^2_t+(h_2)_y+(h_1)_z=0.
$$
Applying the Legendre transform, $u_1=h_1, \ u_2=h_2, \ H=u^1h_1+u^2h_2-h$, one can rewrite these equations in the equivalent form
\begin{equation}
(u_1)_x+(u_2)_z+(H_1)_t=0, ~~~ (u_2)_y+(u_1)_z+(H_2)_t=0.
\label{111}
\end{equation}
Notice that $H(u_1, u_2)$ is defined up to an arbitrary quadratic form (all quadratic terms in $H$ can be eliminated by appropriate linear changes of the independent variables). Our first result is the following

\begin{theorem} Any integrable system (\ref{111}) is necessarily linear (that is,  the potential $H$ is quadratic in $u_1, u_2$).
\end{theorem}

\centerline{\bf Proof:}

\medskip

\noindent Our strategy will be to consider reductions of the system (\ref{111}) to various $(2+1)$-dimensional systems. In fact, it will be sufficient to look at reductions governing  traveling wave solutions. If the original system (\ref{111}) is integrable, all such reductions must be integrable as well.  Since the integrability conditions for $(2+1)$-dimensional two-component systems of hydrodynamic type are explicitly known \cite{1}, this will provide a set of  {\it necessary} conditions for the integrability of the system (\ref{111}). It turns out that these conditions are very strong indeed, leading to the non-existence of non-quadratic integrable potentials $H$.

Setting in the equations (\ref{111}) $\partial_z=\mu \partial_t$, which is equivalent to seeking solutions in the form ${\bf u} (x, y, t+\mu z)$, one obtains a $(2+1)$-dimensional Hamiltonian system
$$
(u_1)_x+(H_1+\mu u_2)_t=0, ~~~ (u_2)_y+(H_2+\mu u_1)_t=0,
$$
with  the  Hamiltonian density $H(u_1, u_2)+\mu u_1 u_2$. According to our philosophy we have to require that it is integrable for an arbitrary value of the parameter $\mu$. The integrability conditions (\ref{H}) readily imply that the corresponding $H$ must be cubic in $u_1, u_2$, and Theorem 1 tells us that the only two  `suspicious' cases to consider are $H=\frac{1}{6} u_2^3$ and $H=\frac{1}{2}u_1u_2^2$ (recall that we  ignore  quadratic terms in $H$). In the first case the system (\ref{111}) takes the form
$$
(u_1)_x+(u_2)_z=0, ~~~ (u_2)_y+(u_1)_z+u_2(u_2)_t=0.
$$
Setting here $x=y$ (this amounts to seeking traveling wave solutions in the form ${\bf u} (x+ y, z, y)$,
one obtains a $(2+1)$-dimensional system
\begin{equation}
(u_1)_x+(u_2)_z=0, ~~~ (u_2)_x+(u_1)_z+u_2(u_2)_t=0.
\label{S1}
\end{equation}
We recall that the paper \cite{1} provides a complete set of the  integrability conditions for two-component hydrodynamic type systems represented in the form
$$
\left(
\begin{array}{c}
v \\
\ \\
w
\end{array}
\right)_t+
\left(
\begin{array}{cc}
a & 0 \\
\ \\
0 & b
\end{array}
\right)
\left(
\begin{array}{c}
v \\
\ \\
w
\end{array}
\right)_x+
\left(
\begin{array}{cc}
p&q \\
\ \\
r&s
\end{array}
\right)
\left(
\begin{array}{c}
v \\
\ \\
w
\end{array}
\right)_y=0.
$$
The integrability conditions constitute a complicated  over-determined  system of PDEs for the coefficients $a, b, p, q, r, s$ as functions of $v, w$. Representing the equations (\ref{S1}) in the form
$$
\left(
\begin{array}{c}
u_1 \\
\ \\
u_2
\end{array}
\right)_x+
\left(
\begin{array}{cc}
0 & 0 \\
\ \\
0 & u_2
\end{array}
\right)
\left(
\begin{array}{c}
u_1 \\
\ \\
u_2
\end{array}
\right)_t+
\left(
\begin{array}{cc}
0&1 \\
\ \\
1&0
\end{array}
\right)
\left(
\begin{array}{c}
u_1 \\
\ \\
u_2
\end{array}
\right)_z=0
$$
one can verify that these integrability  conditions are not satisfied. Thus, the $(3+1)$-dimensional system corresponding to $H=\frac{1}{6} u_2^3$ is not integrable. Similarly, for
 $H=\frac{1}{2}u_1u_2^2$  the system (\ref{111}) takes the form
$$
(u_1)_x+(u_2)_z+u_2(u_2)_t=0, ~~~ (u_2)_y+(u_1)_z+u_2(u_1)_t+u_1(u_2)_t=0.
$$
Setting, again, $x=y$, and changing to the new dependent variables $v=u_1+u_2, \ w=u_2-u_1$, one obtains the system
$$
\left(
\begin{array}{c}
v \\
\ \\
w
\end{array}
\right)_x+
\left(
\begin{array}{cc}
1 & 0 \\
\ \\
0 & -1
\end{array}
\right)
\left(
\begin{array}{c}
v \\
\ \\
w
\end{array}
\right)_z+
\left(
\begin{array}{cc}
\frac{3v+w}{4}&\frac{v-w}{4} \\
\ \\
\frac{v-w}{4}&-\frac{v+3w}{4}
\end{array}
\right)
\left(
\begin{array}{c}
v \\
\ \\
w
\end{array}
\right)_t=0,
$$
which also does not satisfy the integrability conditions. This finishes the proof of Theorem 4.

\medskip

Our next result shows that any three-component  $(3+1)$-dimensional integrable Hamiltonian system associated with a non-singular Poisson bracket of hydrodynamic type is either linear or reducible.  Any such system can be brought to a canonical form
\begin{equation}
u^1_t+(h_1)_x=0, ~~~ u^2_t+(h_2)_y=0, ~~~ u^3_t+(h_3)_z=0,
\label{3+1}
\end{equation}
with the Hamiltonian operator
$$
\left(
\begin{array}{ccc}
d/dx & 0 & 0\\
0 & d/dy & 0\\
0 & 0 & d/dz
\end{array}
\right).
$$
Performing the Legendre transform one obtains
$$
(H_1)_t+ (u_1)_x =0, ~~~ (H_2)_t+(u_2)_y=0, ~~~ (H_3)_t+(u_3)_z=0,
$$
or, in matrix form,
$$
A_0{\bf u}_t+A_1{\bf u}_x+A_2{\bf u}_y+A_3{\bf u}_z=0,
$$
where the $3\times 3$ matrices $A_i$ are given by
$$
A_0=\left(
\begin{array}{ccc}
H_{11} & H_{12} & H_{13}\\
H_{12} & H_{22} & H_{23}\\
H_{13} & H_{23} & H_{33}
\end{array}
\right), ~
A_1=\left(
\begin{array}{ccc}
1 & 0 & 0\\
0 & 0 & 0\\
0 & 0 & 0
\end{array}
\right), ~
A_2=\left(
\begin{array}{ccc}
0 & 0 & 0\\
0 & 1 & 0\\
0 & 0 & 0
\end{array}
\right), ~
A_3=\left(
\begin{array}{ccc}
0 & 0 & 0\\
0 & 0 & 0\\
0 & 0 & 1
\end{array}
\right).
$$

\begin{theorem}
Any integrable $(3+1)$-dimensional Hamiltonian system (\ref{3+1})  is either linear or reducible.
\end{theorem}

\centerline{\bf Proof:}

\medskip

\noindent As a necessary condition for  integrability,  one has to require  the vanishing of the Haantjes tensor
for an arbitrary matrix of the form
$$
(A_0+\lambda A_1+\beta A_2+\gamma A_3)^{-1}(A_0+\tilde \lambda A_1+\tilde \beta A_2+\tilde \gamma A_3),
$$
which is equivalent to  the vanishing of the Haantjes tensor for any matrix
$\Lambda (A_0+\tilde \Lambda)$ where $\Lambda$ and $\tilde \Lambda$ are arbitrary $3\times 3$ constant coefficient
diagonal matrices. Computing the Haantjes tensor end equating to zero coefficients at different monomials in the
diagonal entries of $\Lambda$ and $\tilde \Lambda$, one  obtains that either all third order derivatives
$H_{ijk}$ are identically zero (this corresponds to linear systems), or
$H_{ij}=H_{ik}=0$ for some $i\ne j\ne k$ (this corresponds to the reducible case).

We would like to conclude this section by formulating the following general

\noindent {\bf Conjecture} {\it There exists no non-trivial integrable Hamiltonian systems of hydrodynamic type in $3+1$ dimensions corresponding to a local Poisson bracket of hydrodynamic type and a local Hamiltonian density}.

\section{Concluding remarks}

We have found a broad class of non-trivial potentials leading to integrable Hamiltonian systems of hydrodynamic type in $2+1$ dimensions. There is a number of natural problems arising in this context, in particular:

\noindent --- Describe the structure of the corresponding Hamiltonian hierarchies. The main difficulty here is the non-locality of  higher symmetries/conservation laws.

\noindent --- Construct the associated Hamiltonian hydrodynamic chains. This requires the introduction of a canonical set of non-local variables reducing all higher flows of the hierarchy to infinite-component systems of hydrodynamic type.

\noindent --- Construct dispersive deformations of the examples arising in the classification, especially those with `elliptic'  Lax pairs.

\noindent --- Study the behavior of exact solutions coming from hydrodynamic reductions.

We hope to address some of these questions elsewhere.

\section*{Acknowledgements}

We thank B Dubrovin,  O Mokhov,  A Odesskii,  M Pavlov and A Veselov for clarifying discussions.
This research was supported by the EPSRC grant EP/D036178/1.
The work of EVF was also partially supported by the
European Union through the FP6 Marie Curie RTN project ENIGMA (Contract
number MRTN-CT-2004-5652), and the ESF programme MISGAM.
The work of VVS was  partially supported by the RFBI-grant 08-01-00461. VVS also thanks IHES, where the final part of this research was completed, for their hospitality.

\end{document}